\documentclass[breaklinks, colorlinks, twocolumn]{aastex631}   
\hypersetup{linkcolor=blue,citecolor=blue,filecolor=cyan,urlcolor=magenta}

\usepackage{enumitem}
\usepackage{mathtools}
\usepackage{graphicx}
\usepackage{amssymb}
\usepackage{amsmath}
\usepackage{ulem}
\usepackage{bm}
\usepackage{epstopdf}
\usepackage{color}
\usepackage{subfigure}
\usepackage{cancel}

\font\fiverm=cmr5             \font\sevenrm=cmr7

          \font\sixrm=cmr6       

\def\dover#1#2{\hbox{${{\displaystyle#1 \vphantom{(} }\over{
   \displaystyle #2 \vphantom{(} }}$}}

{\catcode`\@=11                                                  
\gdef\SchlangeUnter#1#2{\lower2pt\vbox{\baselineskip 0pt\lineskip0pt    
\ialign{$\m@th#1\hfil##\hfil$\crcr#2\crcr\sim\crcr}}}}           
\def\gtrsim{\mathrel{\mathpalette\SchlangeUnter>}}               
\def\lesssim{\mathrel{\mathpalette\SchlangeUnter<}}    

\def\fsc{\alpha_{\hbox{\sevenrm f}}}

\def\lambar{\lambda\llap {--}}

\def\teq#1{$\, #1\,$}                         % text equation
\def\rns{R_{\hbox{\sixrm NS}}}
\def\mns{M_{\hbox{\sixrm NS}}}
\def\rS{r_{\hbox{\sixrm S}}} % Schwarzschild radius

\def\erg{\varepsilon}
\def\eesc{\varepsilon_{\rm esc}}
\def\thetakB{\Theta_{\hbox{\sixrm kB}}}

\def\deltaE{\delta_{\hbox{\fiverm E}}}

\def\Thetae{\Theta_{\hbox{\sixrm e}}}
\def\thetaE{\theta_{\hbox{\fiverm E}}}
\def\rEvec{\boldsymbol{r}_{\hbox{\fiverm E}}}
\def\rEvechat{\hat{\boldsymbol{r}}_{\hbox{\fiverm E}}}
\def\kEvechat{\hat{\boldsymbol{k}}_{\hbox{\fiverm E}}}

\def\Omegavechat{\hat{\boldsymbol{\Omega}}}

\def\Bvec{\boldsymbol{B}}

\def\kvec{\boldsymbol{k}}
\def\kvechat{\hat{\boldsymbol{k}}}
\def\rvec{\boldsymbol{r}}

\def\rmax{r_{\hbox{\sevenrm max}}}
\def\emax{\erg_{\hbox{\sevenrm max}}}
\def\PsiE{\Psi_{\hbox{\sixrm E}}}

\def\Omegavec{\boldsymbol{\hat{\Omega}}}

\def\thetaf{\theta_{\hbox{\sixrm f}}}
\def\eescin{\varepsilon_{\rm esc, in}}
\def\eescout{\varepsilon_{\rm esc, out}}
\def\eescside{\varepsilon_{\rm esc, side}}

\shorttitle{Hard X-ray Opacity in Magnetar Magnetospheres}
\shortauthors{Hu et al.}

\begin{document}

%\title{\rm \uppercase{High-Energy Photon Opacity in the Twisted Magnetospheres of Magnetars} }
\title{HIGH-ENERGY PHOTON OPACITY IN THE\\ TWISTED MAGNETOSPHERES OF MAGNETARS}

\correspondingauthor{K. Hu}
\email{kh38@rice.edu}

\author[0000-0002-9705-7948]{Kun Hu}
\affiliation{Department of Physics and Astronomy - MS 108, Rice University, 6100 Main St., Houston, TX 77251-1892, USA}

\author[0000-0003-4433-1365]{Matthew G. Baring}
\affiliation{Department of Physics and Astronomy - MS 108, Rice University, 6100 Main St., Houston, TX 77251-1892, USA}

\author[0000-0001-6119-859X]{Alice K. Harding}
\affiliation{Theoretical Division, Los Alamos National Laboratory, Los Alamos, NM 58545, USA}

\author[0000-0002-9249-0515]{Zorawar Wadiasingh}
\affiliation{Department of Astronomy, University of Maryland, College Park, Maryland 20742, USA}
\affiliation{Astrophysics Science Division, NASA Goddard Space Flight Center, Greenbelt, MD 20771, USA}
\affiliation{Center for Research and Exploration in Space Science and Technology, NASA/GSFC, Greenbelt, Maryland 20771, USA}

\begin{abstract}
Magnetars are neutron stars characterized by strong surface
magnetic fields generally exceeding the quantum critical value of 44.1 TeraGauss.
High-energy photons propagating in their magnetospheres can be attenuated by QED
processes like photon splitting and magnetic pair creation. In this paper, we
compute the opacities due to photon splitting and pair creation by photons emitted
anywhere in the magnetosphere of a magnetar.  Axisymmetric, twisted dipole field
configurations embedded in the Schwarzschild metric are treated.  The paper computes
the maximum energies for photon transparency that permit propagation to infinity in curved spacetime. 
Special emphasis is given to cases where photons are generated along magnetic field
loops and/or in polar regions; these cases directly relate to resonant inverse
Compton scattering models for the hard X-ray emission from magnetars and Comptonized
soft gamma-ray emission from giant flares.  We find that increases in magnetospheric
twists raise or lower photon opacities, depending on both the emission locale, and
the competition between field line straightening and field strength enhancement. 
Consequently, given the implicit spectral transparency of hard X-ray bursts and persistent
``tail'' emission of magnetars, photon splitting considerations constrain their
emission region locales and the twist angle of the magnetosphere; these
constraints can be probed by future soft gamma-ray telescopes such as COSI and
AMEGO. The inclusion of twists generally increases the opaque volume of pair
creation by photons above its threshold, except when photons are emitted in polar
regions and approximately parallel to the field.
\end{abstract}

\vspace{-15pt}

\section{Introduction}
 \label{sec:Intro}

Magnetars are highly-magnetized neutron stars with periods \teq{P} generally in the
2--12 sec range. They exhibit persistent X-ray emission in the \teq{<10}keV band, with
both thermal and non-thermal components \citep[e.g., see][]{Vigano13}, with
luminosities \teq{L_X \sim 10^{33}-10^{35}}erg/sec that exceed the electromagnetic
torque spin-down values \teq{L_{\rm sd}\propto {\dot P}/P^3} inferred from observed
values of \teq{P} and \teq{\dot P}.  
About a third of the population also exhibits persistent, 
hard non-thermal emission in the 10-300 keV band. 
This timing information is employed to discern that
their surface magnetic fields mostly exceed the quantum critical value of \teq{B_{\rm
cr}=m_e^2c^3/(e\hbar)\approx 4.41\times 10^{13}}Gauss, where the electron cyclotron and
rest mass energies are equal. Such superstrong fields are a distinguishing hallmark of
magnetars: they are believed to power their sporadic X-ray burst emission \citep{DT92,TD96}.  
Most of this transient activity consists of short hard X-ray flares of subsecond duration with
luminosities in the \teq{10^{38}}erg/sec \teq{ < L_X < 10^{42}} erg/sec range. The
trapping of magnetospheric plasma for such durations requires the presence of strong
magnetic fields.  For recent magnetar reviews, see \cite{Turolla15} and \cite{KB17}. 
The observational status quo of magnetars is also summarized in the McGill Magnetar
Catalog \citep{OK14} and its online
portal\footnote{http://www.physics.mcgill.ca/~pulsar/magnetar/main.html}.

While the thermal signals from magnetars below around 5 keV provide key information for
understanding their surfaces, it is the magnetospheric signals above 10 keV that are
germane to the study here.  There are three types of hard X-ray/soft gamma-ray emission
exhibited by magnetars.  The first of these consists of steady, hard, non-thermal pulsed
spectral tails that have been detected in around ten magnetars
\citep{Kuiper06,Gotz06,Hartog08a,Hartog08b,Enoto-2010-ApJ,Younes17}. These luminous
tails, usually fit with power-law spectral models of \teq{F_{\nu}} index between 0 and
1, extend up to 150 - 250 keV, with a turnover around 500 - 750 keV implied by
constraining pre-2000 COMPTEL upper limits.   The {\it
Fermi}-Gamma-Ray Burst Monitor (GBM) has also observed these tails \citep{terBeek12},
providing the sensitivity to better measure the flux above 100 keV. Over the last
decade, much more {\it Fermi}-GBM data on magnetars has been accumulated, enabling
improved spectral definition, yet with no material change in the overall shape of the
tails (Kuiper, private comm.).  The {\it Fermi}-Large Area Telescope (LAT)
has not detected this component \citep[e.g.,][]{Abdo-2010-ApJL,Li-2017-ApJ}.  
The most prominent model for the generation of these
tails is inverse Compton scattering, resonant at the cyclotron frequency in the strong
magnetar fields: see \cite{BH07,FT07,Beloborodov13,Wadiasingh18}.

Transient, recurrent bursts are observed for many magnetars, both for the soft gamma
repeater (SGR) and anomalous X-ray pulsar (AXP) varieties. Some episodes of such bursts
last hours to days, over which tens to hundreds of individual short bursts can occur. 
Given their typical \teq{\sim 0.01-0.3}sec durations and super-Eddington luminosities,
they must be generated from highly optically thick magnetospheric regions.  The bursts
mostly have emission below around 100 keV, with a spectral breadth
\citep{Gogus99,Feroci04,Israel08,Lin12,vdH12,Younes14} that indicates thermal gradients
and strong Comptonization \citep{Lin-2011-ApJ} in the emission regions.   For occasional
exceptional bursts, the observed maximum observed energy is somewhat higher, for example, 
two anomalously hard bursts from SGR 1900+14 \citep{Woods-1999-ApJ} observed in late
1998 and early 1999.  A more recent notable exemplar is the steeper spectrum FRB-X burst
\citep{Mereghetti-2020-ApJ,Li-2021-NatAst,Ridnaia-2021-Nat-Ast} associated with the fast
radio burst (FRB) seen in April 2020 from SGR 1935+2154, being spectrally unique among
the population of bursts detected from this magnetar \cite{Younes-2021-Nat-Ast}.

There are also the rare giant flares from magnetars, highly optically thick to Compton
scattering and with luminosities \teq{10^{44} - 10^{47}} erg/sec at hard X-ray energies
extending up to around 1 MeV.  They have been observed for only two SGRs in the Milky
Way  \citep[e.g.,][]{Hurley99,Hurley-2005-Nature} and one in the Large Magellanic Cloud 
\citep[e.g.,][]{Mazets79}.  They are characterized by a short, intense spike of duration
\teq{\sim 0.2}s, followed by a pulsating tail lasting several
minutes that is spectrally softer.  In April 2020, a giant flare was detected from a
magnetar in the galaxy NGC 253 at 3.5 Mpc distance, exhibiting only the initial
spike in {\it Fermi}-GBM observations up to around 3 MeV \citep{Roberts-2021-Nature},
followed by delayed GeV emission seen by {\it Fermi}-LAT \citep{Ajello-2021-Nat-Ast}. 
This event provided the clearest view to date of the MeV-band spectral evolution of
giant flares.

Key questions surrounding these three varieties of magnetar hard X-ray signals are:
where in the magnetosphere do they originate, how does the magnetic field modulate
and power their activity, and what physics controls their spectral character?   A
central element concerns how prolifically electron-positron pairs are created.  This
paper focuses on two exotic QED processes operating in magnetar magnetospheres, magnetic
pair creation \teq{\gamma\to e^{\pm}}, permitted above its threshold energy of
\teq{2m_ec^2}, and magnetic photon splitting \teq{\gamma\to\gamma\gamma}, which can
attenuate photons all the way down to around 50 keV in magnetar fields \citep{BH01}.
These processes are permitted only in the presence of magnetic fields
\citep[e.g.,][]{Erber66,Adler71}, and both possess rates that are extremely sensitive to
the strength of the magnetic field {\bf B}, the angle \teq{\thetakB} that photons
propagate relative to the field direction, and the energy of a photon.  Many of 
their pertinent properties including magnetospheric opacity have been addressed in
numerous papers \citep{Baring95,HBG97,BH98,BH01,SB14}. In particular, the
suppression of pair creation by photon splitting in neutron star magnetospheres was
discussed by \cite{BH98} as a possible means for generating a ``radio death line" for
high-field pulsars.

Recently, \cite{Hu-2019-MNRAS} calculated photon splitting and pair creation
opacities in the inner magnetospheres of high {\bf B} neutron stars, applicable to
arbitrary colatitudes and a substantial range of altitudes in closed field line zones.
This work determined both attenuation lengths and escape
energies for each process, the latter being the maximum photon energy for which the
magnetosphere is transparent. Yet it restricted its focus to general relativistic dipole
field geometries. 

A major element of the magnetar paradigm is the force-free MHD
distortion of field line morphology incurred by large pair currents
\citep[e.g.,][]{TLK02,Beloborodov13,CB17}. 
In this scenario, magnetic and other stresses in the crust are released 
via surface shear motions that rotate the external field lines within flux tubes/surfaces,
thereby generating magnetospheric helicity. The twisted configuration can be 
sustained for extended time periods by magnetospheric currents due to \teq{e^--e^+} 
pairs created in electrostatic potential gaps, likely somewhat near polar zones.
The twists require both toroidal and poloidal currents \citep[e.g.,][]{TLK02, Beloborodov-2009}:
 toroidal currents straighten 
the poloidal magnetic field lines, and poloidal currents support the toroidal field components.
These twisted fields, superposed on the global quasi-dipolar magnetic structure, serve as
energy reservoirs and thus couple intimately to magnetospheric activation. 
Twists thus increase the local magnetic field strength \teq{B},
change the radii of field curvature, two influences that profoundly alter photon
splitting and pair opacity in magnetars, raising and (sometimes) lowering the escape energy.  Here we
extend the analysis of \cite{Hu-2019-MNRAS} to address twisted field morphologies,
embedding the ideal MHD flat spacetime prescription in the Schwarzschild metric.

This construction is detailed in Section~\ref{sec:twist_field}, following a summary of
the opacity geometry and the QED physics of magnetic photon splitting and pair creation
in Section~\ref{sec:setup}.  Results are presented in Sections~\ref{sec:escape_energies}
and~\ref{sec:opacity_volume}.  The introduction of twisted fields to the magnetosphere 
increases the volumes of opaque regions spanning mid-latitudes to the magnetic equator, 
since the twist changes the field morphology and increases the field magnitude. 
For photons emitted from specific field loops, twisted magnetospheres establish higher 
photon escape energies because of the straightening of the field lines and the 
accompanying rise of the emission altitudes.  The effectiveness of photon splitting 
as a competitor to pair creation increases near the polar regions
as the cumulative azimuthal shear angle \teq{\Delta \phi_{\rm tw}} of the magnetic twist
increases.  Contextual discussion of the results is presented in Sec.~\ref{sec:discussion}.

\section{Opacity Geometry and Physics}
 \label{sec:setup} 

This section summarizes the set-up for our radiative transfer 
opacity calculations, namely the geometry and the QED physics pertinent to 
the attenuation.  

\subsection{Radiative Transfer Geometry}
 \label{sec:geometry}
 
The optical depth for a photon emitted at any locale \teq{\rEvec} 
in the magnetosphere (in the observer's 
coordinate frame; OF)  in direction \teq{\kEvechat} is given by 
\begin{equation}
   \tau(l)  \;=\; \int_0^l {\cal{R}}  
   \bigl(\erg_{\infty}, \kEvechat , \rEvec ; \thetakB, \Bvec \bigr)
   \,ds \quad ,   
 \label{eq:opt_depth}
\end{equation}
where \teq{\cal{R}} is the attenuation coefficient (in units of cm$^{-1}$) and 
\teq{l} is the cumulative proper length of the photon trajectory. 
Here \teq{\erg_{\infty} m_ec^2} is the observed photon energy at infinity.
In considering general relativistic photon propagation, the optical depth is integrated 
over the geodesic path-length \teq{s}, sampling different local inertial frames (LIFs) with altitude.
Thus \teq{\Bvec\equiv \Bvec_{\rm GR}} is the LIF frame magnetic field, specified in
Eq.~(\ref{eq:Bvec_GR}) below, and \teq{\thetakB} is the angle between the field and 
the photon momentum in the LIF, which satisfies Eq.~(\ref{eq:thetakB_def}).
The attenuation length \teq{L} is defined \citep{BH01,SB14,Hu-2019-MNRAS} to be the 
proper distance over which optical depth \teq{\tau(L)=1}.   The principal quantity of interest 
in our opacity considerations is the escape energy \teq{\eesc}, which is defined as 
the OF photon energy at which the attenuation length \teq{L} becomes infinite:
\begin{equation}
   1 \;=\; \int_0^{\infty} {\cal{R}}  
   \bigl(\eesc, \kEvechat , \rEvec ; \thetakB, \Bvec \bigr)
   \,ds \quad .
 \label{eq:eesc_def}
\end{equation}
Since the attenuation coefficients increase rapidly with photon energy for both photon splitting and pair creation, 
the optical depth satisfies \teq{\tau<1} for photons with energies \teq{\erg} smaller than the escape energy \teq{\eesc}.
Thus the escape energy \teq{\eesc} represents the photon energy below which the magnetosphere is transparent. 
It is a strong function of the emission locale (at which \teq{s=0}) and the magnetic field morphology,
features that are central to the considerations of the paper.

\begin{figure}
\vspace*{-5pt}
\centerline{\hskip -0pt\includegraphics[width=8.3cm]{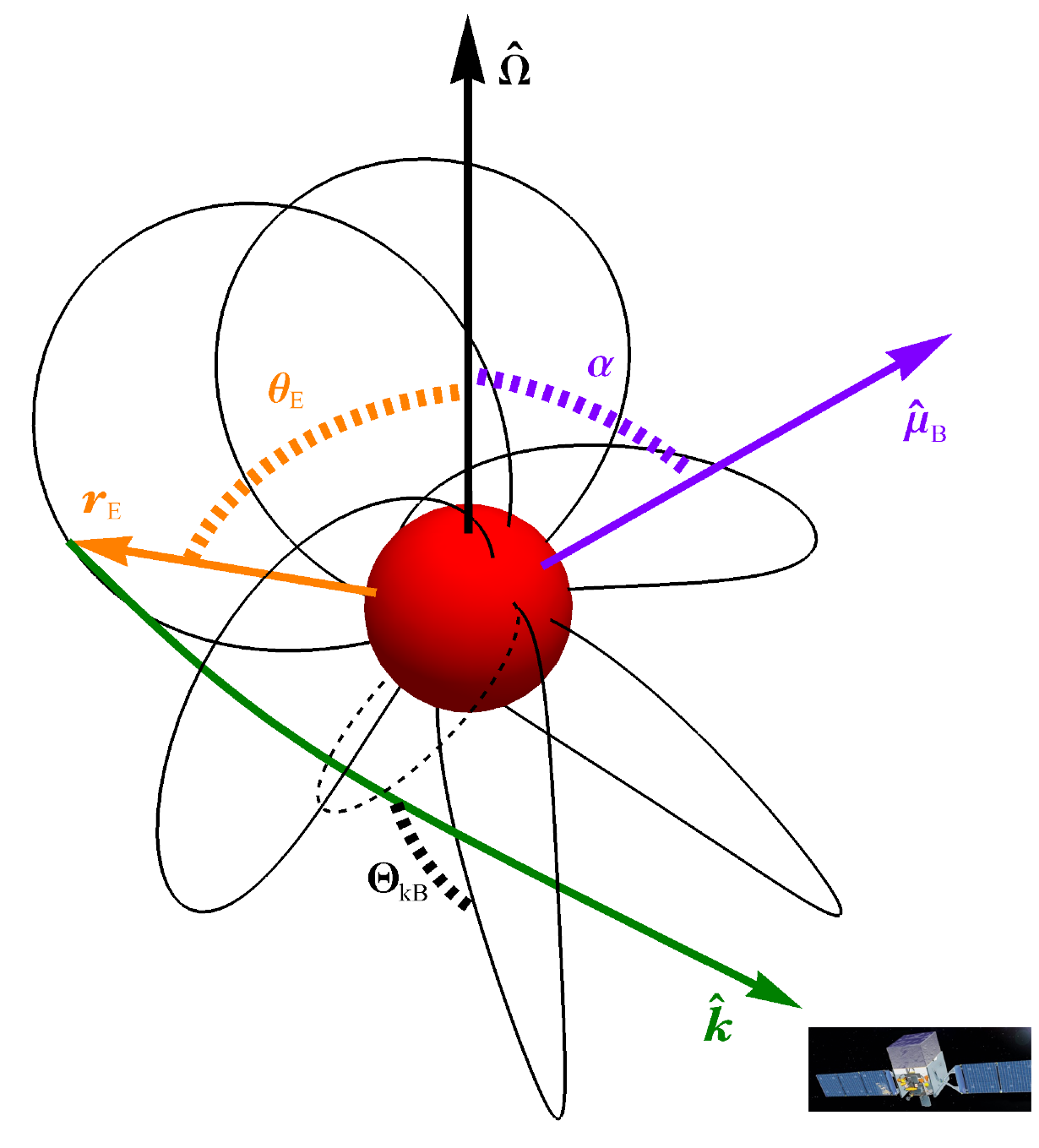}}
\vspace*{-5pt}
\caption{
Schematic diagram displaying the photon propagation geometry in a twisted magnetosphere. 
The rotation axis \teq{\Omegavechat} is depicted as black arrow, and the 
magnetic dipole moment 
vector \teq{\hat{\boldsymbol \mu}_{\hbox{\sixrm B}}} at one particular rotational phase is 
displayed as the purple arrow.  The star's magnetic inclination is \teq{\alpha =\arccos 
[ \hat{\boldsymbol \mu}_{\hbox{\sixrm B}} \cdot \Omegavechat ]}.
One photon is emitted at \teq{\rEvec} (emission polar angle
\teq{\thetaE=\arccos [ \rEvechat \cdot \Omegavechat ]}) and propagates 
through the foreground of the magnetosphere: 
the trajectory of the photon is represented by a green curve, lying in front of the star;
\teq{\kvechat} signifies the changing direction (unit momentum vector) of the photon.  
Twisted field loops with a 
\teq{p=0.75} ($\Delta \Phi_{\rm tw}=84^{\circ}$) parameterization in the 
Schwarzschild metric (\teq{\mns = 1.44M_{\odot}})
are plotted as black curves anchored at a colatitude of \teq{30^\circ} relative to the magnetic axis. 
An untwisted field loop anchored at the same magnetic colatitude is also depicted as a 
dashed black curve for comparison.  A representative angle \teq{\thetakB} between 
the photon trajectory and a local magnetic field is highlighted.
 \label{fig:atten_geom}
 \vspace{-9pt}
 }
\end{figure}

The geometry of the attenuation process in the twisted magnetosphere of a rotating 
neutron star is depicted in Fig.~\ref{fig:atten_geom}, with the {\sl Fermi} 
Gamma-Ray Space Telescope image being from its mission web page.\footnote{\tt https://fermi.gsfc.nasa.gov}   
The inclination angle between the 
stellar rotation axis and the magnetic axis is denoted as \teq{\alpha}, and the colatitude 
of the emission locale relative to the rotation axis is denoted as \teq{\thetaE}.
We consider photon propagation in the curved spacetime described by the 
Schwarzschild metric, fixing the stellar radius at \teq{\rns = 10^6} cm and 
stellar mass at \teq{\mns = 1.44M_{\odot}} throughout.   For the schematic in 
Fig.~\ref{fig:atten_geom}, the photon path starts with emission parallel to \teq{\Bvec}, 
the focal case of Section~\ref{sec:escape_energies_parallel}.  The optical depth in 
Eq.~(\ref{eq:opt_depth}) is then integrated numerically over the geodesic path-length 
\teq{s}, and at each point along the path, \teq{{\cal R}} is specified in the LIF.  
The trajectories of photons are integrated numerically in our calculation. 
These protocols are detailed in Section 5.1 of \cite{Hu-2019-MNRAS}.
The opacity is very sensitive to the angle between the external magnetic field and 
the momentum vector of the propagating photon, which is given by
\begin{equation}
   \cos\thetakB \; =\;  \dover{\kvechat\cdot\boldsymbol{B}}{\vert \boldsymbol{B}\vert} 
   \quad \hbox{or}\quad
   \sin\thetakB \; =\;  \dover{\vert \kvechat\times\boldsymbol{B}\vert}{\vert \boldsymbol{B}\vert} \quad .
 \label{eq:thetakB_def}
\end{equation}
Although our formalism and associated computations are presented in curved spacetime, 
flat spacetime results can be easily obtained by specializing to \teq{\mns\rightarrow 0},
and are used as a check of the results (see below).

In all results in this paper, the modulation of viewing geometry with stellar
rotation phase will not be considered, as it was in Section 4.3 of
\cite{Hu-2019-MNRAS}, where the pertinent behavior was detailed sufficiently. The
primary interest here is in how twisted field morphology influences escape energies
and the volume of magnetospheric opacity. Accordingly, it suffices to consider
aligned rotators with \teq{\alpha=0} in the ensuing exposition, so that
\teq{\thetaE} represents the magnetic colatitude of emission at position vector \teq{\rEvec}.

\subsection{Photon Splitting and Pair Creation}
 \label{sec:splitting_pair_physics}
 
The attenuation of photons is calculated for two linear polarization modes, 
namely \teq{\parallel} (ordinary) mode and \teq{\perp} (extraordinary) mode.
Here \teq{\perp} and \teq{\parallel} refer to the states where photon electric vector 
is locally perpendicular and parallel to the plane containing the photon momentum 
vector \teq{\kvec} and the external field \teq{\Bvec}, respectively. 
These two linear modes approximately represent the polarization eigenmodes 
of soft X rays when vacuum polarization dominates the dielectric tensor of the 
magnetosphere of a magnetar.  As the photon propagates in the magnetosphere, 
its electric field vector evolves adiabatically following the change of the magnetic 
field direction, with the birefringent QED vacuum ensuring that the polarization 
state (i.e., \teq{\perp} or \teq{\parallel}) remains unchanged during propagation 
\citep[see][]{Heyl00}.  
This adiabatic polarization evolution persists out to the 
polarization-limiting radius, which is mostly beyond the escape altitudes 
for X-rays and gamma-rays that are highlighted in the various figures below. 
This feature simplifies the polarization transport considerably.

The opacity is computed for the two key processes that can be prolific 
in strong-field QED, namely photon splitting and pair creation.
Here we provide a summary of the essentials of the rates for these 
processes, and detailed discussions of the physics for them can be found in 
\cite{HBG97}, \cite{BH01} and also in \cite{HL06}.
 
Magnetic photon splitting is a third-order QED process in which a single photon splits into two lower-energy photons.
The CP invariance of QED permits only three splitting channels, namely:
\teq{\perp \rightarrow \parallel \parallel}, \teq{\parallel \rightarrow \perp \parallel} 
and \teq{\perp \rightarrow \perp \perp} \citep[see][]{Adler71}.
In the domain of weak vacuum dispersion, \teq{\perp \rightarrow \parallel \parallel} 
is the only CP-permitted channel that satisfies energy-momentum conservation \citep{Adler71}. 
This kinematic selection rule may not hold when plasma dispersion competes 
with the vacuum contribution in lower fields at higher altitudes where the cyclotron 
resonance can become influential.  Thus, in the following presentation, we will consider 
all the CP-permitted splitting channels, and focus especially on polarization-averaged 
results, which differ only modestly from results where only 
\teq{\perp \rightarrow \parallel \parallel} is activated.

In the low-energy limit well below the pair creation threshold, applying 
to hard X-rays that are of principal interest here,
the reaction rates for all the CP-permitted modes can be expressed as 
\citep{BH01}
\begin{eqnarray}
   {\cal R}^{\rm sp}_{\perp\to\parallel\parallel} &=& 
   \dover{\fsc^3}{60\pi^2\lambar_c}\, \erg^5\, B^6\, {\cal M}_1^2 \, \sin^6\thetakB
   \; =\; \dover{1}{2}\, {\cal R}^{\rm sp}_{\parallel\to\perp\parallel} \quad ,
  \nonumber\\[-5.5pt]
  &&  \label{eq:splitt_atten} \\[-5.5pt]
   {\cal R}^{\rm sp}_{\perp\to\perp\perp} &=& 
   \dover{\fsc^3}{60\pi^2\lambar_c }\, \erg^5\, B^6\, {\cal M}_2^2 \, \sin^6\thetakB \quad .\nonumber
\end{eqnarray}
Here \teq{\fsc} is the fine structure constant, \teq{\lambar_c} is the reduced 
Compton wavelength of electron, and the field strength \teq{B} is expressed 
in the unit of the quantum critical field \teq{B_{\rm cr}}.  The LIF photon energy \teq{\erg} is 
scaled by the rest mass energy of electron.  For unpolarized photons, the attenuation coefficient 
can be obtained averaging Eq.~(\ref{eq:splitt_atten}):
\begin{equation}
   {\cal R}^{\rm sp}_{\rm ave} \; =\;  \dover{\fsc^3}{120\pi^2\lambar_c }\, \erg^5\, B^6\,
    \Bigl( 3{\cal M}^2_1+{\cal M}^2_2\Bigr) \, \sin^6\thetakB \quad ,
 \label{eq:split_pol_ave}
\end{equation}
a result detailed in \cite{Hu-2019-MNRAS}.  The reaction rate coefficients 
\teq{{\cal M}_1, {\cal M}_2} derived from squares of the matrix elements, are 
purely functions of \teq{B} in this low energy \teq{\erg\ll 1} domain, possessing the forms
\begin{equation}
   {\cal M}_{\sigma} \; =\; \dover{1}{B^4}\int_{0}^{\infty} \dover{ds}{s}e^{-s/B} \, \Lambda_{\sigma} (s) \quad ,
 \label{eq:calM_i_form}
\end{equation}
with 
\begin{eqnarray}
   \Lambda_1(s) & = & 
   \left(-\frac{3}{4s}+\frac{s}{6}\right)\frac{\cosh s}{\sinh s}+\frac{3+2s^2}{12\sinh^2s}+\frac{s\cosh s}{2\sinh^3s} \quad , \nonumber\\[-5.5pt]
 \label{eq:Lambda_s_def}\\[-5.5pt]
   \Lambda_2(s) & = & 
   \frac{3}{4s}\frac{\cosh s}{\sinh s}+\frac{3-4s^2}{4\sinh^2s}-\frac{3s^2}{2\sinh^4s} \quad .\nonumber
\end{eqnarray}
For low fields \teq{B\ll 1}, \teq{{\cal M}_1\approx 26/315} and \teq{{\cal M}_2\approx 48/315}  
are independent of \teq{B}, but at highly supercritical fields \teq{B\gg 1} possess 
\teq{{\cal M}_1\propto B^{-3}} and \teq{{\cal M}_2\propto B^{-4}} dependences.
 
One-photon pair creation is a first-order QED process that is allowed in 
the presence of a strong external field because momentum conservation 
orthogonal to {\bf B} is then not operable.  This conversion process is 
extremely efficient when \teq{B\gtrsim 0.1} and can only proceed when 
the photon energy is above the \teq{\erg_{\perp} \equiv \erg \sin\thetakB=2} threshold. 
The produced electrons occupy excited Landau levels in the external magnetic field. 
The attenuation coefficient exhibits a sawtooth structure since it 
diverges at the threshold of each Landau level accessed
\citep[see][for detailed calculations]{DH83, BK07}.
The attenuation coefficient for pair creation can be expressed in the form of
\begin{equation}
   {\cal R}^{\rm pp}_{\perp,\parallel} \; =\; \dover{\fsc}{\lambar_c} 
   \, B\sin\thetakB\,  {\cal F}_{\perp,\parallel} \left(\erg_\perp,\, B\right)
  \label{eq:pp_general}
\end{equation}
for the \teq{\perp, \parallel} photon linear polarizations.  Since the 
forms we employ for the \teq{{\cal F}_{\perp,\parallel}} functions have been 
presented in several other papers and are somewhat lengthy, 
they are listed in Appendix A.

\section{Axisymmetric Twisted Magnetic Fields in a Schwarzschild Metric}
 \label{sec:twist_field}
 
The strong sensitivity of the photon splitting and pair creation rates to the 
angle \teq{\thetakB} between the photon momentum and the local field 
indicates that opacity to each process will be enhanced 
significantly when the radius of curvature of the field is decreased.  
Thus we anticipate that the introduction of twists to a quasi-dipolar field morphology, 
including toroidal components (\teq{B_{\phi}}), will increase these opacities
for certain emission locales and directions, to be identified.
Twisted field morphology is a principal means of storing extra energy in 
the magnetosphere for subsequent dissipation \citep{WL92}.  
To assess the nature and general level of the impact of field twists on 
splitting and pair conversion rates, it is appropriate to adopt a relatively 
simple twist configuration.  The most convenient prescription is a 
``self-similar'' axi-symmetric form that was originally derived 
in the context of the solar corona \citep[e.g.,][]{WL92,Wolfson95}; it is 
a power-law radial solution of the ideal MHD equations.  Just as this
idealized form does not precisely model solar coronal field loops, 
we contend that it will not describe the active magnetar field flux tubes 
that can be inferred from soft X-ray data \citep{Younes-2022-ApJL}.

In the application of axisymmetric twists to magnetars, the twisted field 
components can be expressed in flat spacetime as \citep{TLK02, Pavan09}
\begin{eqnarray}
     && \Bvec_f \; \equiv\; \bigl(B_{r f}, B_{\theta f}, B_{\phi f} \bigr) \nonumber \\[-5.5pt]
  \label{eq:b_twist_fl}\\[-5.5pt]
     && = \frac{B_p}{2}\left(\frac{\rns}{r}\right)^{p+2}\left[-\frac{dF}{d\mu}, 
        \frac{pF}{\sin{\theta}},\sqrt{\frac{C\,p}{p+1}}\frac{F^{1+1/p}}{\sin{\theta}}\right], \nonumber
\end{eqnarray}
where \teq{F=F(\mu)} is the magnetic flux function that satisfies the 
Grad-Shafranov equation \citep{Lust54,Grad-1958-UNconf,Shafranov-1966-RvPP}
\begin{equation}
    p(p+1)F(\mu)+(1-\mu^2)\frac{d^2 F(\mu)}{d \mu^2}=-CF^{1+2/p}(\mu).
 \label{eqn:G_S_eqn}
\end{equation}
Here \teq{C} and \teq{p} are constants and \teq{\mu=\cos{\theta}} represents 
the magnetic colatitude.  The domain of interest is \teq{0 \leq p \leq 1},
though we note that \teq{p >1} cases (with different boundary conditions) 
could treat multipole field 
components that generally store magnetic energy on scales smaller than the 
twisted regions explored here -- consideration of photon opacity in multipolar field 
configurations is deferred to future work.

Eq.~(\ref{eqn:G_S_eqn}) is non-linear unless \teq{C=0}, 
in which case the field configuration is purely poloidal (\teq{B_{\phi}=0}).
The boundary conditions can be specified using the symmetry: 
\teq{B_{r}(\mu=0)=0\rightarrow F'(0)=0};\teq{B_{\phi}(\mu=1)=0\rightarrow F(1)=0}. 
The third condition can be specified by either fixing the polar field strength 
\citep[see][]{TLK02} or fixing the magnetic flux threading each hemisphere 
\citep{Wolfson95}. In our analysis, a fixed polar field strength is more 
appropriate, so we choose \teq{B_{r}(\mu=1)=B_{p}\rightarrow F'(1)=-2} 
as the third boundary condition, closing the system.
The field configuration collapses to a magnetic dipole 
with \teq{F(\mu )=1-\mu^2} when \teq{p \rightarrow 1}, while it becomes a 
split monopole with  \teq{F(\mu )=2\vert 1-\mu \vert} when \teq{p \rightarrow 0} 
\citep[see][]{Wolfson95, Pavan09}. In these cases, we have \teq{C(0)=0=C(1)}.  

For a specific \teq{p} value, the solution of Eq.~(\ref{eqn:G_S_eqn}) 
and the constant \teq{C} can be determined using the previous boundary 
conditions. In particular, we numerically solve Eq.~(\ref{eqn:G_S_eqn}) 
using a shooting method combined with the 4th-order Runge-Kutta technique. 
For a given \teq{p} value, we choose a test \teq{C=C_{\rm test}}, for 
\teq{0<C_{\rm test}<1}.  Then Eq.~(\ref{eqn:G_S_eqn}) can be solved 
for \teq{F(\mu)} and \teq{F'(\mu)} starting from \teq{F(1)=0} and \teq{F'(1)=-2} 
with the Runge-Kutta technique.  This gives us a \teq{F'(0)} value for this 
specific \teq{C_{\rm test}}. Then we keep varying the \teq{C_{\rm test}} value 
and redo the Runge-Kutta process until the boundary condition
\teq{F'(0) = 0} is met. Thereafter, \teq{F(\mu)}, \teq{F'(\mu)} and \teq{C} 
are solved for the give \teq{p} value. 
For our opacity computations, the solutions of \teq{F(\mu)} and  \teq{F'(\mu)} 
are then tabulated for each \teq{p}, and the tables are used 
to construct the field structure in either Eq.~(\ref{eq:b_twist_fl}) or 
Eq.~(\ref{eq:Bvec_GR}) at any point in the magnetosphere.

Given our specific choice of boundary conditions, we can integrate the 
Grad-Shafranov equation using successive integration by parts on the 
derivative term:
\begin{equation}
  \int_0^1 (1-\mu^2) \, \dover{d^2F(\mu )}{d\mu^2} \, d\mu
   \;=\;  - 2 \int_0^1 F(\mu ) \, d\mu\quad .
 \label{eq:G-S-deriv_term}
\end{equation}
It then follows that an integral form of the Grad-Shafranov equation 
for our boundary value problem is
\begin{eqnarray}
  (p-1) (p+&&2) \int_0^1 F(\mu ) \, d\mu   \nonumber\\[-5.5pt]
     \label{eq:G-S-integrated} \\[-5.5pt]
   && +\; C(p)\,\int_0^1 \bigl[ F(\mu ) \bigr]^{1+2/p} \, d\mu  \; =\; 0  \;\; .\nonumber
\end{eqnarray}
We used this form to provide consistency checks on the numerical 
determinations of \teq{F(\mu )} and \teq{C(p)} for each particular \teq{p}.
The result was that both were accurate to within around 1\% or better
for all \teq{p} values.

The \teq{p} value controls the cumulative angular/toroidal shear along a specific field loop.
The shear angle is an integration over the \teq{B_{\phi}/B_{\theta}} ratio, 
from the footpoint at magnetic colatitude \teq{\theta} to the point of maximum 
altitude at the equator.  This can be expressed as \citep{Wolfson95,TLK02}
\begin{equation}
    \Delta \phi (\mu) = 2\!\! \int^{\mu}_{0} \frac{B_{\phi}}{B_{\theta}}\frac{d\mu}{1-\mu^2} 
    =2\sqrt{\frac{C(p)}{p(1+p)}}\int^{\mu}_{0}\frac{F^{1/p} d\mu}{1-\mu^2}\; ,
 \label{eq:del_phi}
\end{equation}
where \teq{\arccos \mu} is the magnetic colatitude of the footpoint.  The factor 
of 2 accounts for the contribution to the shear from both hemispheres. 
For a field loop anchored near the magnetic poles, the maximal shear (or twist) angle 
is \teq{\Delta\phi_{\rm tw} \equiv\Delta \phi (\mu = 1)}
\citep[called \teq{ \Delta \phi_{\rm N-S}} by][]{TLK02}, 
and this decreases monotonically with increasing \teq{p}.
Thus \teq{\Delta\phi_{\rm tw}} serves as an alternate parameter 
of common usage for the axisymmetric twist solution.

We graphically compared our numerical solution of Eq.~(\ref{eqn:G_S_eqn}) with 
prior results, finding that the \teq{C-p} and \teq{p-\Delta \phi_{\rm tw}} relations 
from our G-S solver code were in excellent agreement (better than 1\%) with the depictions 
in the upper right of Fig.~2 of \cite{Pavan09} and Fig.~2 of \cite{TLK02}, respectively.   
Our \teq{F(\mu )} also agrees very well with those for \teq{p=0.02, 0.5, 1.0} displayed 
in Fig.~2 of \cite{Pavan09}.  Yet, our \teq{F(\mu)} function deviates by around 6\% 
for the \teq{p=0.2, 0.7} examples in Fig.~2 of \cite{Pavan09}; the origin of this discrepancy 
is unknown.  We checked our results using Eq.~(\ref{eq:G-S-integrated}) and found 
very good agreement (better than 0.1\% for \teq{p>0.1}), in particular validating our
\teq{p=0.2, 0.7} solutions.  These tests underpin our confidence that our encoded 
flat spacetime Grad-Shafranov solver operates correctly. 

\begin{figure*}
\vspace*{10pt}
\centerline{
\includegraphics[width=18.0cm]{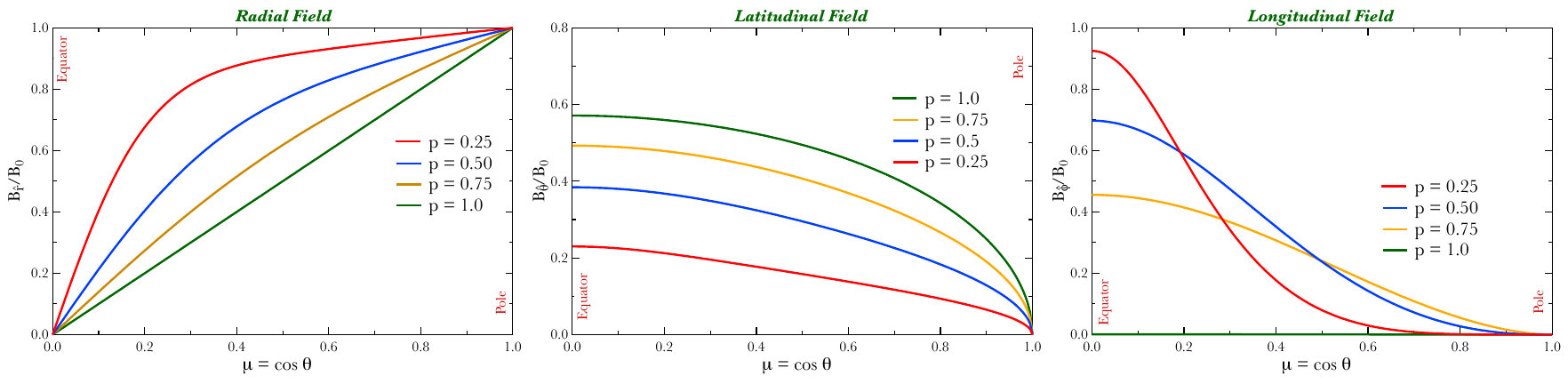}}
\vspace*{-5pt}
\caption{
The \teq{B_{\hat r}}, \teq{B_{\hat \theta}}, and \teq{B_{\hat \phi}} components 
(from left to right) of the magnetic field in the Schwarzschild metric's LIF at the surface of 
a \teq{1.44M_{\odot}} neutron star of radius \teq{\rns=10^6}cm. They are plotted as 
functions of the magnetic colatitude \teq{\mu = \cos{\theta}}
with \teq{p=0.25, 0.5, 0.75}, and 1.0 (dark green, pure dipole case),
ordered from top to bottom as indicated in the legends. 
The field component magnitudes are normalized by the field strength at the 
magnetic pole in the LIF, namely \teq{B_0 \approx 1.488 B_p}.
 }
  \label{fig:B_r_theta_phi_twist}
\end{figure*}

One comparatively simple way to combine information from axisymmetric 
solutions of ideal MHD with general relativity is to directly embed the field components 
determined above in a magnetic field framework appropriate for the Schwarzschild
metric, such as that developed in \cite{Petterson-1974-PhRvD,Wasserman83,Muslimov-1986-SvA} for dipole fields.  
In that dipolar configuration, the curved spacetime modifies the radial (\teq{B_r}) and
polar (\teq{B_{\theta}}) field components by different analytic factors.  These 
factors are described for the Schwarzschild geometry by the functions
\begin{eqnarray}
   \xi_r (\Psi) & \; =\;  & - \frac{3}{\Psi^3} \left[ \log_e(1-\Psi) + \Psi + \frac{\Psi^2}{2} \right] \nonumber\\[-5.5pt]
 \label{eq:xi_r_theta_def}\\[-2.5pt]
   \xi_{\theta} (\Psi) & \; =\; & \frac{6}{\Psi^3\sqrt{1-\Psi}} \left[ (1-\Psi)\, \log_e(1-\Psi) + \Psi - \frac{\Psi^2}{2} \right] ,\nonumber
\end{eqnarray}
where \teq{\Psi=\rS/r} for a Schwarzschild radius \teq{\rS=2GM/c^2}.  In the 
flat spacetime domain, \teq{\xi_r \rightarrow 1}, \teq{\xi_{\theta} \rightarrow 1} 
when \teq{\Psi \rightarrow 0}.  Note that these definitions of \teq{\xi_r}
and \teq{\xi_{\theta}} are scaled differently from those listed in \cite{Hu-2019-MNRAS}
and references therein.  

The GR modification to the field components in the azimuthal and polar angle 
directions is identical, i.e., \teq{\xi_{\phi} \equiv \xi_{\theta}}, an intuitive choice.  
The Schwarzschild geometry is invariant under 
rotations about the radial direction, and specifically so for 90 degree rotations 
that affect the coordinate mappings \teq{\theta \leftrightarrow \pm\phi}.  One deduces 
that the GR component amplification factors \teq{\xi_{\phi}} and \teq{\xi_{\theta}}
are equal for any field configuration.  For a mathematical 
determination of this identity for arbitrary multipoles, the reader 
can inspect Eqs.~(5) and~(6) of \cite{Muslimov-1986-SvA}.
Observe that the common enhancement factor \teq{\xi_{\theta}} 
cancels out when calculating the shear angle, so that Eq.~(\ref{eq:del_phi}) is still 
valid for our GR implementation.  This symmetry does not extend 
to the Kerr metric.

The flat spacetime field components in Eq.~(\ref{eqn:G_S_eqn}) 
are then embedded in the Schwarzschild metric to generate 
\teq{\Bvec_{\rm GR} = (B_{ {\hat r}}, \, B_{ {\hat \theta}}, \, B_{ {\hat \phi}})},
the LIF frame magnetic field (using the hat notation for LIF coordinates).  Thus, 
\begin{equation}
    \Bvec_{\rm GR} \; =\;  \Bigl[\xi_r(\Psi)B_{r f}, \; \xi_{\theta}(\Psi)B_{\theta f}, 
       \; \xi_{\theta}(\Psi) B_{\phi f} \Bigr] \quad ,
 \label{eq:Bvec_GR}
\end{equation}
where \teq{B_{r f}}, \teq{B_{\theta f}} and \teq{B_{\phi f}} are the field components 
in flat spacetime, as specified in Eq.~(\ref{eq:b_twist_fl}). 
This is the magnetic field form employed in the opacity calculations of this paper.
Specific field loops can be computed numerically for a particular 
field line footpoint on the surface, or a maximum altitude \teq{r_{\rm max}}, 
using parametric equations derived from the field components in the 
Schwarzschild metric:
\begin{equation}
   \dover{dr}{d\theta} \; = \; r\sqrt{1-\Psi} \, \frac{B_{ {\hat r}}}{B_{ {\hat \theta}}}
   \quad \hbox{and} \quad
   \dover{d\phi}{d\theta} \; =\; \frac{1}{\sin{\theta}} \frac{B_{ {\hat \phi}}}{B_{ {\hat \theta}}} \quad .
 \label{eq:field_loci}
\end{equation}
This protocol was adopted for the visual depiction of the twisted field lines 
in Fig.~\ref{fig:atten_geom}, which were determined using \teq{\Psi \approx 0.425} 
for \teq{M=1.44M_{\odot}} and \teq{R=10^6}cm.

Fig.~\ref{fig:B_r_theta_phi_twist} displays the different components 
of curved spacetime twisted magnetic field at the stellar surface. 
The field strengths are scaled by the zero-twist polar field strength \teq{B_0 = \xi_r(\Psi )\, B_p} 
in the Schwarzschild metric, for \teq{\Psi \approx 0.425}.
For a fixed colatitude, the radial \teq{B_{\hat r}} component increases and the 
latitudinal (polar) \teq{B_{\hat \theta}} component decreases with decreasing \teq{p}. 
The longitudinal (azimuthal) \teq{B_{\hat \phi}} component is more concentrated at \teq{\mu=0} with smaller \teq{p}.
When \teq{p} is small, the radial component dominates the magnetic field and asymptotically forms a split monopole. 
For magnetospheric regions beyond the stellar surface, all the components are 
simply multiplied by a factor of \teq{(\rns/r)^{p+2}}, yet subject to different GR corrections,
leaving the relative apportionment between the field components only mildly altered.

As an alternative approach, \cite{Kojima17} directly solved the 
Grad-Shafranov equation in the Schwarzschild metric,
assuming an axisymmetric magnetosphere and a power-law form for the flux function.
The GR metric modifications appear as modifiers to the various gradient operators
embedded in the field's vector potential, and a Legendre polynomial expansion 
in the \teq{\mu =\cos\theta} angular portion leads to radial eigenfunctions 
of hypergeometric function form.  This protocol is more involved numerically than 
our choice here, and perhaps not as simple to digest.  Yet it too generates 
an idealized solution through the parameterized power law assumption in specifying the toroidal 
\teq{B_{\hat\phi}} component.  In reality, the expected restriction of zones of enhanced 
magnetization to flux tubes implies a caveat 
to all axisymmetric MHD models of neutron star magnetospheres.
We adopt our protocol for its conceptual and numerical simplicity.

\section{Escape Energies}
 \label{sec:escape_energies}
 
In this section, the character of X-ray and gamma-ray escape energies is presented,
for photons emitted at the stellar surface and in a magnetosphere described by the
twisted field paradigm.  After summarizing the opacity validation protocols, we
consider three cases for emphasis.  The first is where photons are emitted parallel
to the local field direction; this is directly connected to the resonant inverse
Compton scattering model of the persistent hard X-ray emission of magnetars.  The
next case of interest is for emission perpendicular to the local field, which
approximately represents conditions most likely to be sampled in the bursts that are
a common occurrence for magnetars.  Finally, the focus turns to polar regions where
the field line radii of curvature are large; this constitutes a case that may be
quite relevant to the initial spikes of magnetar giant flares.

Our numerical calculations are validated by several methods.  First, our photon
trajectory integration is tested by comparing the integrated trajectories to the
asymptotic trajectory formula presented in \cite{Poutanen-2020-AA}. The deviation of
our computed light bending angles with that formula is around 0.003\% for the 
worst case scenario where the photon is emitted horizontally from the stellar surface,
i.e., at periastron.  We compared the magnetic field
strength and direction at select points along photon trajectories with analytic
values for the cases with \teq{p= 0} and \teq{1}, yielding good agreement. 
The attenuation coefficient during the propagation of photons was plotted 
(not shown) as functions of pathlength and values checked at select positions 
along trajectories with the employed analytic forms for LIF values of the photon 
energy, the field strength and \teq{\thetakB}.  Finally, calculated escape energies 
and opaque zones in the \teq{p\rightarrow1} domain (green curves in
Figures.~\ref{fig:B_r_theta_phi_twist_b100} and~\ref{fig:B_r_theta_phi_twist_b10})
are in excellent agreement with the dipole results in Figures 5, 6, 9, and 10 in
\cite{Hu-2019-MNRAS}, which also serve as checks of the opacity calculation.

\subsection{Photons Emitted Parallel to {\bf B}}
 \label{sec:escape_energies_parallel} 

The first case to be considered is when the photons are emitted parallel to the 
magnetic field at the point of emission.  For magnetars, this closely matches 
the expectations of the popular resonant inverse Compton scattering (RICS) 
models for the production of their persistent hard X-ray emission above 
10 keV \citep{BH07,FT07,Beloborodov13,Wadiasingh18}, discussed below.
It is also relevant to curvature radiation emission components from high-field pulsars 
such as PSR B1509-58 \citep{HBG97}, if they are generated by primary electrons 
accelerated in polar cap or slot gap potentials with \teq{\boldsymbol{E} \cdot \boldsymbol{B}
\neq 0} in the inner magnetosphere \citep{DH96,MH04}.

\begin{figure*}
\vspace*{10pt}
\centerline{
\includegraphics[width=17.5cm]{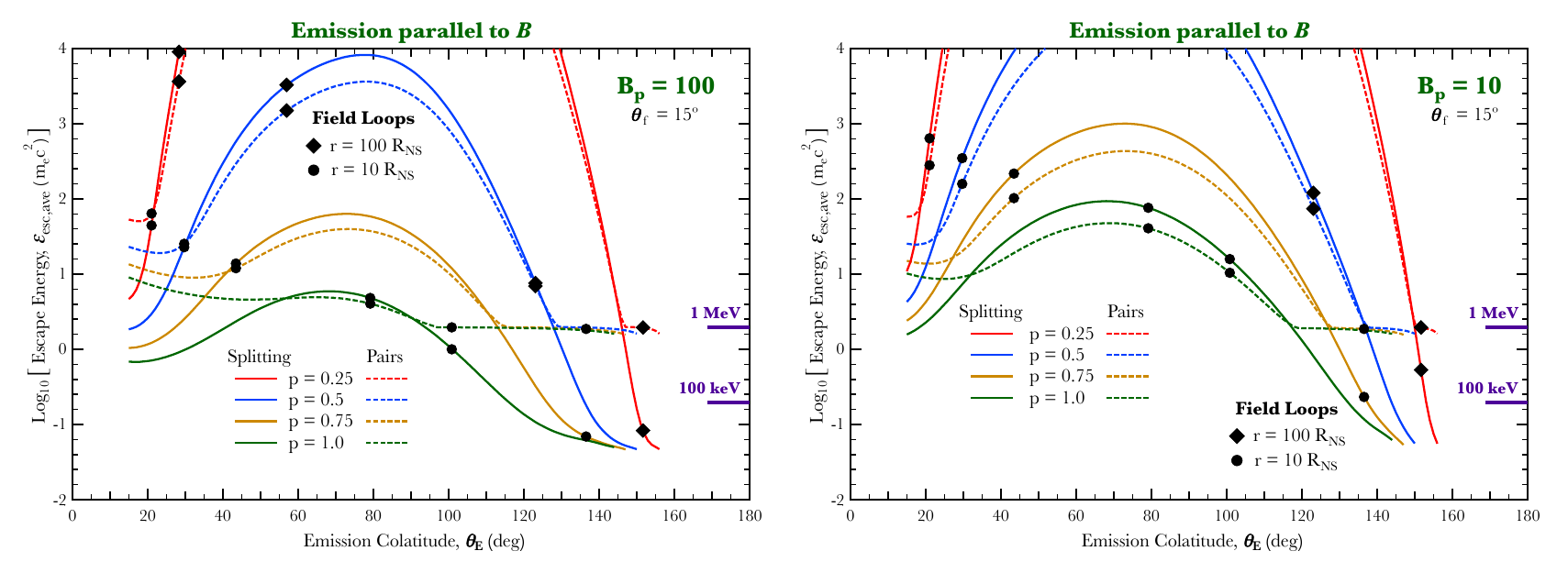}}
\vspace*{-5pt}
\caption{
Polarization-averaged escape energies for photon splitting (solid curves) or pair creation 
(dashed curves) for photons emitted at points along twisted magnetic field loops with 
\teq{p=0.25,\;0.5,\;0.75} and \teq{1.0}, ordered from top to bottom as indicated 
in the panel legends.   The surface polar fields are \teq{B_p=100} and \teq{B_p=10} for the 
left and right panels, respectively.  The footpoint colatitude \teq{\thetaf} of the field loops is fixed 
at \teq{15^\circ}, and photons start their trajectories parallel to the local field (\teq{\Theta_e=0^{\circ}}). 
The black circles and diamonds label the colatitudes where the emission altitude equals 
\teq{10 \rns} and \teq{100 \rns}, respectively. 
Short horizontal lines on the right-hand side of each panel indicate the energies of 1 MeV and 100 keV.
}
  \label{fig:eesc_indiv_spli_pair}
\end{figure*}
\begin{figure*}
\vspace*{10pt}
\centerline{
\includegraphics[width=17.5cm]{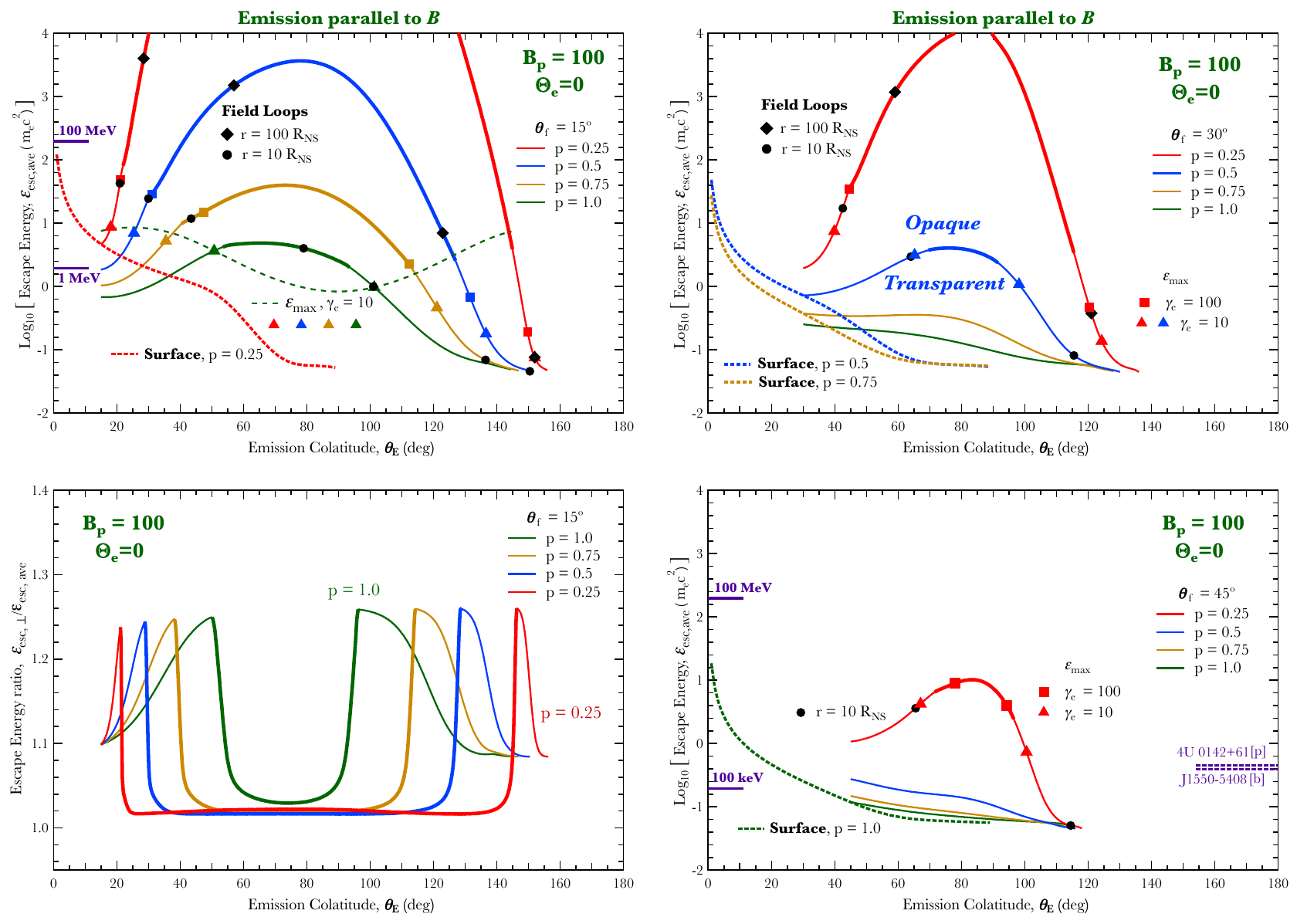}}
\vspace*{-5pt}
\caption{
Escape energies for combined photon splitting and pair creation for photons emitted from the
stellar surface (dashed curves) or along twisted magnetic field loops (solid curves)
with \teq{B_p = 100}, and for \teq{p=0.25,\;0.5,\;0.75} and \teq{1.0}, generally ordered from 
top to bottom (from outside to inside on the lower left) as indicated in the panel legends.
Emission is parallel to the local field (\teq{\Theta_e=0^{\circ}}).  The heavyweight
portion of the curves marks portions of the field lines where pair creation
dominates photon splitting in determining the escape energy.  The top and
right panels depict the polarization-averaged escape energies for field loops with
footpoint colatitudes \teq{\theta_{f} = 15^{\circ}}, \teq{30^{\circ}} and
\teq{45^{\circ}} respectively. The bottom left panel displays the ratio of
\teq{\perp} mode escape energy \teq{\varepsilon_{\rm esc,\perp}} to the
polarization-averaged escape energy \teq{\varepsilon_{\rm esc,ave}} for
\teq{\theta_{f} = 15^{\circ}} that is depicted at the upper left.  The colored triangles
(\teq{\gamma_e = 10}) and squares (\teq{\gamma_e = 100}) represent the emission
colatitude where the escape energy is equal to the maximum energy \teq{\emax}
produced by RICS (see Eq.~\ref{eq:emax} and associated text).  The dashed green
curve in the \teq{\theta_{f} = 15^{\circ}} panel (upper left) defines the RICS
\teq{\emax} for the general relativistic dipole configuration (\teq{p=1}). The filled black circles and
diamonds label the locales on the field loops corresponding to an altitude equal to
\teq{10\rns} and \teq{100\rns}, respectively. In the lower right panel are marker
energies (purple dashed lines) indicating the approximate maximum observed energy in
two magnetars with polar fields somewhat close to the illustrated values, SGR
J1550-5408 (bursts, [b]) and AXPs 4U 0142+61 (persistent emission, [p]); 
see \cite{Hu-2019-MNRAS} 
for details concerning these observational indicators. 
}
  \label{fig:B_r_theta_phi_twist_b100}
\end{figure*}

\begin{figure*}
\vspace*{10pt}
\centerline{
\includegraphics[width=17.5cm]{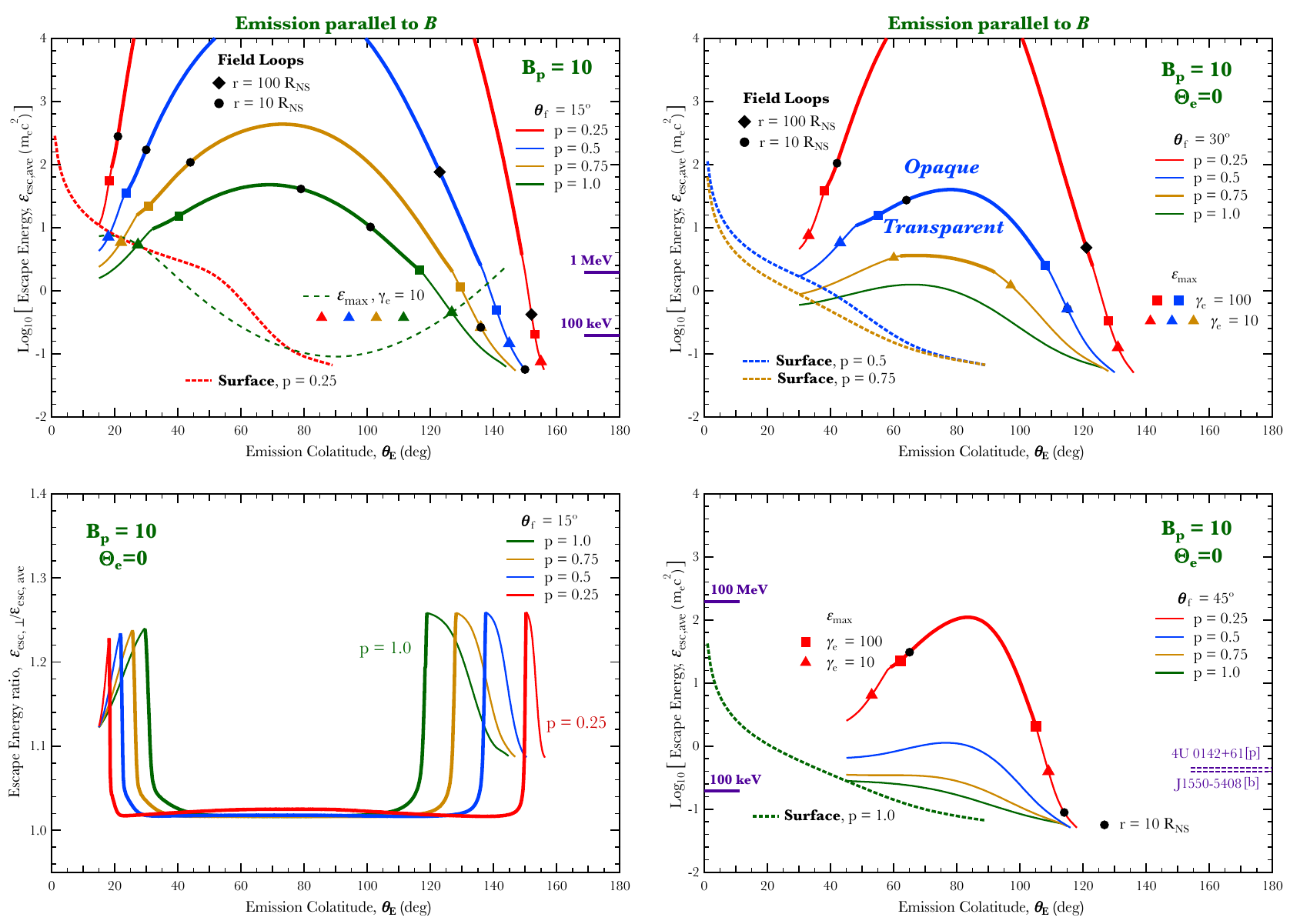}}
\vspace*{-5pt}
\caption{
Escape energies for photon splitting and pair creation for photons emitted from the 
stellar surface (dashed curves) or along twisted magnetic field loops (solid curves).
Again, these are for \teq{p=0.25,\;0.5,\;0.75} and \teq{1.0}, being ordered generally from 
top to bottom (or from outside to inside on the lower left) as indicated in the panel legends.
The surface polar field 
is now \teq{B_p=10}, and emission is again parallel to the local field (\teq{\Theta_e=0^{\circ}}). 
Labels for points in all panels are identical to those in Fig.~\ref{fig:B_r_theta_phi_twist_b100},
as are the observational data markers in the lower right panel.}
  \label{fig:B_r_theta_phi_twist_b10}
\end{figure*}

The opacity for a polarized photon emitted at a specific locale can be obtained
using Eq.~(\ref{eq:opt_depth}) with the attenuation coefficients of photon splitting
and pair creation specified by Eqs.~(\ref{eq:split_pol_ave}) and
(\ref{eq:pp_general}), respectively. Then the escape energy \teq{\eesc} is determined
by adjusting the photon energy so that the cumulative optical depth in propagating
to infinity equals unity, effectively a root solving algorithm for
Eq.~(\ref{eq:eesc_def}).  Thus, \teq{\eesc} depends on the initial
conditions, namely the emission position and direction, and subsequently also on the
resultant path of the photon. Fig.~{\ref{fig:eesc_indiv_spli_pair}} illustrates the
escape energies \teq{\eesc} of photon splitting (solid) and pair creation (dashed)
as functions of the emission colatitudes \teq{\thetaE} for photons originating on
specific field loops, with their initial photon momentum parallel to the local
field, i.e. \teq{\boldsymbol{k} \parallel \boldsymbol{B}}. Escape energies for
combined photon splitting and pair creation with surface polar fields \teq{B_p =
100} and \teq{B_p = 10} are illustrated in Figs.~\ref{fig:B_r_theta_phi_twist_b100}
and \ref{fig:B_r_theta_phi_twist_b10}, respectively. The escape energy loci in both
of these Figures were determined for field loops with one of the footpoint
colatitudes located at \teq{\thetaf=15^\circ}, and for four different twist
parameter \teq{p} values. On each curve in Fig.~{\ref{fig:eesc_indiv_spli_pair}},
the black circles and diamonds mark the colatitudes where the emission locales have
radii of 10 and 100 \teq{\rns} respectively. Decreasing the \teq{p} value enlarges
the field loop so that its maximum altitude is larger (see Table~\ref{table:thetaf_rmax}), 
corresponding to much larger radii of field curvature on average. 

\begin{deluxetable}{|c |c  ccc|cc|cc|cclcc}[htb!]
\centerwidetable
\movetabledown = 20mm
\tablecolumns{13}
\tablecaption{
\teq{\thetaf-\rmax} relations for different \teq{p} values
}
\label{table:thetaf_rmax}
\tablehead{ \multicolumn{1}{c}{}   & \multicolumn{4}{c}{\teq{\rmax \;(\rns)}} \\
\hline
 \multicolumn{1}{|c|}{\teq{p}}  & \multicolumn{1}{c}{\teq{\thetaf = 15^{\circ}}}  & \multicolumn{1}{c}{\teq{\thetaf = 30^{\circ}}} & \multicolumn{1}{c}{\teq{\thetaf = 45^{\circ}}} & \multicolumn{1}{c|}{\teq{\thetaf = 60^{\circ}}} 
}
\startdata
1.00 &10.37	 & 3.01 & 1.67 & 1.22  \\
0.75 &29.92	 & 5.29 & 2.22 & 1.38  \\
0.50 &266.5	 & 18.12 & 4.30 & 1.84 \\
0.25 &2.13\teq{\times10^5}  & 909.6 & 42.48 & 5.69 \\
\enddata
\end{deluxetable}
\vspace{-30pt}

The field loop escape energies in Fig.~{\ref{fig:eesc_indiv_spli_pair}} are 
maximized at colatitudes somewhat smaller than \teq{90^\circ}, and resemble the
bell-shaped behavior presented in Fig.~10 of \cite{Hu-2019-MNRAS} (\teq{p=1} case
only). This is expected because of the high altitudes of emission at
quasi-equatorial colatitudes, where the magnetic fields are relatively weak. Pair
creation dominates the attenuation and determines the escape energy at colatitudes
remote from the footpoints, and these domains are marked as heavyweight portions of
the curves in Fig.~\ref{fig:B_r_theta_phi_twist_b100} (upper left); the lower average fields along the 
photon trajectories tend to favor the dominance of pair conversion over photon splitting
once the energy threshold is exceeded \citep[e.g.,][]{BH01}. 
While the pair conversion escape energies (for an observer at infinity) in 
Fig.~{\ref{fig:eesc_indiv_spli_pair}} generally exceed the threshold of
\teq{2m_ec^2} for pair creation, there are noticeable portions near the highest 
emission colatitudes where values slightly lower than \teq{2m_ec^2} are apparent.
These correspond to inward emission cases where the LIF frame photon energy 
actually exceeds the pair threshold for the inner portion (near periastron) of 
the trajectory to infinity.  For both processes, the escape energy
curves are asymmetric about the equator (\teq{\thetaE=90^\circ}), since photons
emitted from upper and lower hemispheres possess different trajectories. Photons
emitted outward travel to regions of lower field magnitudes and larger radii of
curvature. In contrast, those emitted inward sample fields that are stronger and
field lines that are more highly curved (see the sample trajectories 
in Fig.~\ref{fig:opaque_volume}), conditions conducive to greater opacity, 
thereby lowering \teq{\eesc}.

The key feature of the impact of twists to the magnetospheric geometry is that the
escape energy for loop emission increases as \teq{p} decreases. This is caused by
two effects. At small colatitudes, decreasing \teq{p} enhances the radial
\teq{B_{\hat r}} component. Thus the initial photon momentum \teq{\boldsymbol{k}} is
typical more radial than it is in a dipole field, and \teq{\thetakB} increases more
slowly for lower \teq{p} due to the straighter field lines, unless the photon is
very far from the stellar surface. For moderate colatitudes, decreasing \teq{p}
raises the emission locale to a higher altitude (see the black circles and diamonds
in Fig.~\ref{fig:B_r_theta_phi_twist_b100}), which also reduces the opacity as the
typical value of \teq{\vert \boldsymbol{B}\vert} along the trajectory is lower. 
Accordingly, the higher the axisymmetric MHD twist, the more transparent the
magnetosphere becomes to splitting and pair creation for photons emitted 
parallel to \teq{\Bvec} along field lines.  As \teq{p} declines, the opacities should
rise with the increase of the field magnitude \teq{\vert\Bvec\vert}  that is evident in
Fig.~\ref{fig:B_r_theta_phi_twist}.  Yet, this influence is actually dominated by
the associated increase in the radii of field line curvature for higher twists, which 
reduces the magnetospheric opacity as the field morphology progresses towards that
of a split monopole.

Escape energies for photons emitted from the stellar surface are also depicted in
Fig.~\ref{fig:B_r_theta_phi_twist_b100} as dashed curves. These curves start at
\teq{\thetaE=0^\circ} and decrease with the increase of \teq{\thetaE} because the
field line curvature radii are smaller at non-polar colatitudes. All the curves are
truncated at \teq{\thetaE=90^\circ}, beyond which the emitted photons propagate
inward and therefore never enter the magnetosphere. The escape energy curves for the
loop emission touch the surface curves at the footpoint colatitudes \teq{\thetaf},
where the field loops are anchored at the surface. The escape energies for surface
emitted photons also increase with the decrease of \teq{p}, observing that the
initial photon momentum \teq{\boldsymbol{k}} is more radially directed when \teq{p}
is small.

The resonant upscattering of the surface thermal emission by relativistic
\teq{e^+/e^-} is very efficient in the magnetospheres of magnetars, because the
resonance at the cyclotron frequency (\teq{\varepsilon=B}) increases the cross
section by around 2-3 orders of magnitude above the classical Thomson value when
\teq{B\lesssim1} \citep[e.g.,][]{Gonthier00}. The velocity of the ultra-relativistic
\teq{e^+/e^-} \citep{BH07,Wadiasingh18} is essentially parallel to the magnetic
field direction due to rapid cyclotron cooling perpendicular to the field, modulo
small drift velocity components due to the slow magnetar rotation. Thus the momentum
of the upscattered photon is nearly parallel to the magnetic field as a consequence of
Doppler beaming. This also applies to other emission mechanisms, including curvature
radiation and synchrotron emission for $\gamma$-ray pulsars. The scattered
\teq{e^+/e^-} are assumed to stay in the ground Landau states, which is relevant to
situations where the scattering samples energy near or below the fundamental
cyclotron resonance. This is generally valid due to rapid cooling of \teq{e^+/e^-}
near the cyclotron resonance, which can normally prevent \teq{e^+/e^-} from
encountering higher cyclotron harmonics; see \cite{BWG11} for a discussion.

For this case of scatterings being dominated by those sampling the cyclotron
frequency, the maximum scattered photon energy \teq{\emax} is realized
for a head-on collision with a photon scattering angle of \teq{180^\circ} in
the electron rest frame, which leads to \citep{BWG11} 
\begin{equation}
\vspace{-5pt}
   \emax\;=\;\gamma_e(1+\beta_e) \left(\frac{B}{1+2B}\right)\quad.
 \label{eq:emax}
\end{equation}
Here \teq{\gamma_e} is the Lorentz factor of the \teq{e^+/e^-}, \teq{\beta_e} is 
the ratio of lepton speed to \teq{c}, and \teq{B} is the field strength at the scattering locale.  
This form was employed in \cite{Hu-2019-MNRAS} in Fig.~8 for the dipole case.
For fixed lepton velocity, the maximum scattered photon energy \teq{\emax} solely 
depends on the field strength, and obviously saturates at \teq{\gamma_e(1+\beta_e)/2} 
when \teq{B \gg1}. 

The gravitational-redshifted maximum scattered photon energy
\teq{\emax/(\sqrt{1-\Psi})\approx 1.32 \emax} (for \teq{\Psi = 0.425})
along the \teq{p=1} field loop is displayed as a
sparsely dashed green curve in the upper left panel of
Fig.~\ref{fig:B_r_theta_phi_twist_b100}.  This is specifically for a fixed
\teq{\gamma_e=10}, a value that naturally emerges as a result of rampant RICS
cooling \citep[][]{BWG11,Wadiasingh18} in magnetar-strength fields. The \teq{\emax} trace is
symmetric about the magnetic equator (\teq{\thetaE=90^\circ}), and it intersects the
escape energy curve at \teq{\thetaE\approx 50^\circ, 100^\circ}, which are labeled
by green triangle markers. At low/high colatitudes outside these two intersections,
the maximum photon energy \teq{\emax} is larger than the escape energy \teq{\eesc},
therefore photon splitting will attenuate the highest energies of hard X-ray photons
produced by RICS. The \teq{\emax} curves for field loops with smaller \teq{p} values
share similar behavior and so are not displayed in the figures. The \teq{\emax}
values decline along the trace as the colatitude moves away from the poles towards
the equator because the emission locales are then farther from the surface and the
field strength is weaker. Intersections of loci for \teq{\emax} and \teq{\eesc} for
loops with different \teq{p} values are labeled in the figures with triangle markers
(\teq{\gamma_e=10}) and square markers (\teq{\gamma_e=100}) in the three panels 
that depict escape energies.  The separation of the two intersection points 
increases with a decrease of \teq{p}, so that when \teq{p} is small, photon splitting is 
more likely to attenuate photons at energies below \teq{\emax} near the field line 
footpoints.

The bottom-left panel of Fig.~\ref{fig:B_r_theta_phi_twist_b100} illustrates the
ratio of the \teq{\perp} mode (\teq{\perp\rightarrow \parallel+\parallel},
\teq{\perp\rightarrow e^+e^-}) escape energy to the polarization-averaged escape
energy. The ratios increase with emission colatitude \teq{\thetaE} at small
\teq{\thetaE}; and sharply drop to around unity when pair creation dominates the
opacities (the heavyweight portion demarcates pair conversion dominance for
\teq{\varepsilon_{\rm esc,\perp}}) at quasi-equatorial colatitudes. For the
quasi-polar domains where photon splitting determines the escape energy, the ratio
can be intuitively estimated as \teq{\varepsilon_{\rm esc,\perp}/\varepsilon_{\rm
esc,ave}\approx \left[2{\cal M}_1^2/(3{\cal M}_1^2+{\cal M}_2^2)\right]^{-1/5}}.
This approaches the value \teq{(338/1083)^{-1/5}\approx1.26} in the sub-critical
field domain, which is in agreement with the ratio cusps apparent in
Fig.~\ref{fig:B_r_theta_phi_twist_b100}. For pair creation considerations, the
escape energy ratio can be estimated assuming \teq{3 \exp{[-8/(3 B \varepsilon_{\rm
esc, ave})}]/2\approx\exp{[-8/(3 B \varepsilon_{\rm esc, \perp})}]} in the
sub-critical field domain.  This estimate is obtained using the asymptotic expressions 
for the polarization-dependent and polarization-averaged pair creation rates in \cite{Erber66}, 
which can also be deduced by taking the \teq{\erg_{\perp}\gg 2} limit of the expressions 
in Eq.~(\ref{eq:BK07_perp_asymp}) in Appendix A.  This relation yields \teq{\varepsilon_{\rm
esc,\perp}/\varepsilon_{\rm esc,ave}\approx 1+3 B \eesc \ln{(3/2)}/8\approx 1+
0.152B \eesc}, which is very close to unity for high altitude emission near the
equator. Since these ratios are not vastly different from unity, it is sufficient to
employ just polarization-averaged opacity determinations for deriving the
representative character of opacity in the ensuing exposition.

Fig.~\ref{fig:B_r_theta_phi_twist_b10} is a \teq{B_p=10} analogue of the escape
energy results in Fig.~\ref{fig:B_r_theta_phi_twist_b100}.  This value is close to
the surface field strengths of most magnetars. Since the attenuation of both photon
splitting and pair creation is positively correlated with the field strength, the
escape energies are generally higher than those in
Fig.~\ref{fig:B_r_theta_phi_twist_b100}. So the domination of pair creation
(weighted curves) covers a larger portion of the escape energy curves. Yet the
general shapes of the escape energy curves are quite similar to those in
Fig.~\ref{fig:B_r_theta_phi_twist_b100}, a consequence of the employment of an
identical selection of field morphologies.

\vspace{20pt}

\subsection{Photons Emitted Perpendicular to {\bf B}}

Modest or large emission angles \teq{\Thetae} to the local field are expected for
magnetar bursts and flares, where radiation comes from highly optically-thick
``fireballs'' along magnetic flux tubes populated by quasi-thermalized
electron-positron pair plasma. To accommodate this situation and provide a contrast
to results from Section \ref{sec:escape_energies_parallel}, we explore the escape
energy for photons emitted initially perpendicular to the local \teq{\Bvec} field,
which might be the primary direction for radiation to escape from the fireball in
the closed-field region. Here we consider four different azimuthal directions for
such \teq{\Thetae = 90^{\circ}} cases, namely primarily outward emission
(\teq{\kvec\propto\Bvec\times\hat{\phi}} so that \teq{\kvec \cdot \hat{r} >0}),
inward emission (\teq{\kvec\propto\hat{\phi}\times\Bvec} with \teq{\kvec \cdot
\hat{r} <0}) both in the \teq{r-\theta} planes, and two sideways emission cases
[\teq{\pm \kvec\propto(\Bvec\times\hat{\phi})\times\Bvec}] that are essentially
orthogonal to active magnetic flux tubes.

\begin{figure}
\vspace*{0pt}
\centerline{\hskip 0pt\includegraphics[width=8.5cm]{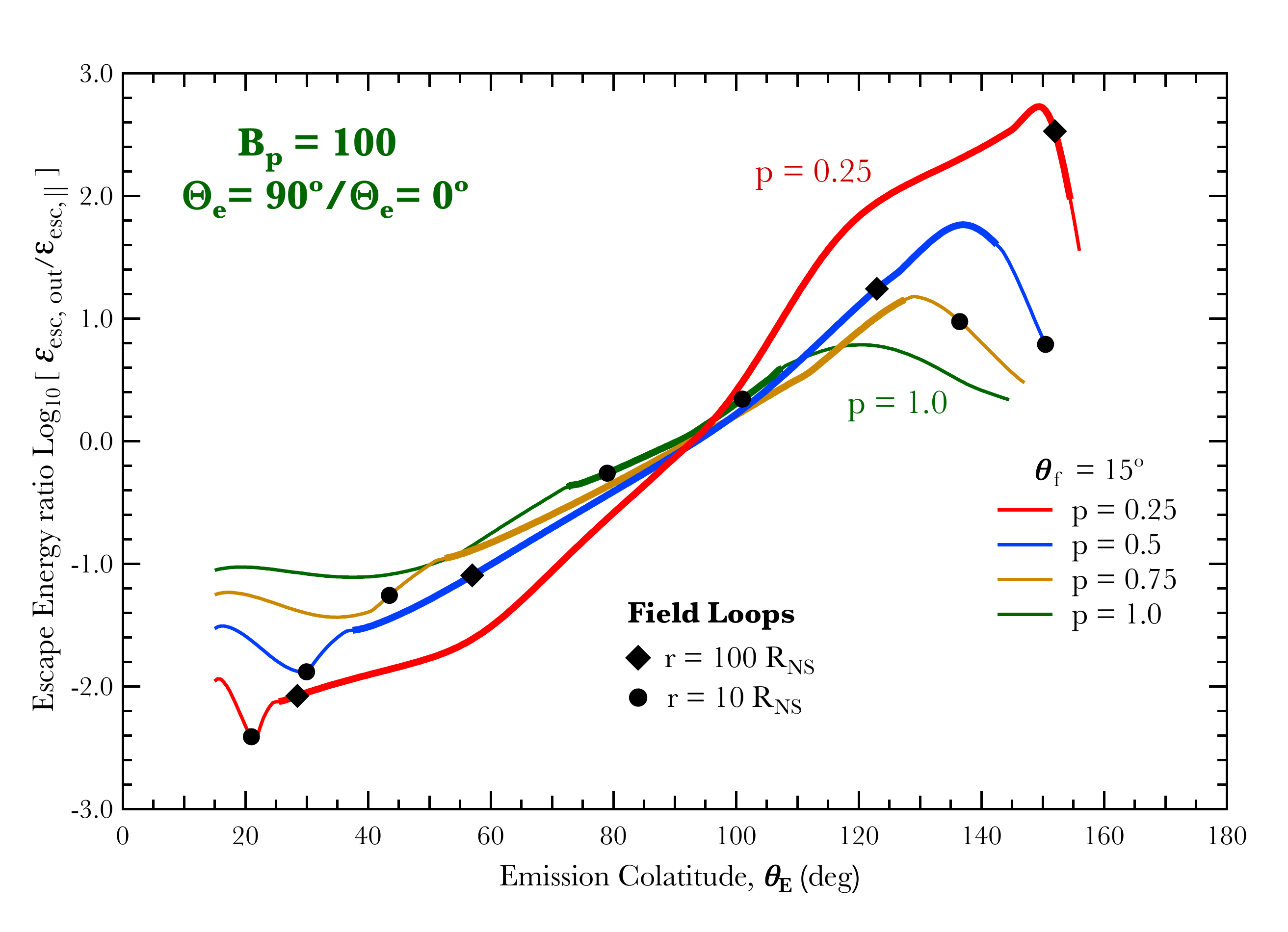}}
\vspace*{-10pt}
\caption{
The ratio of the escape energies (logarithmic scale) of photons emitted perpendicular 
to the local {\bf B} direction (\teq{\Theta_{\rm e}=90^{\circ}}) to those emitted parallel 
to the field (\teq{\Theta_{\rm e}=0^{\circ}}).  These ratios are plotted as functions 
of the emission colatitude on field loops with a fixed footpoint colatitude of 
\teq{\thetaf = 15^\circ}, for four different \teq{p} values, as indicated. 
For the perpendicular emission, the initial photon momentum 
satisfied \teq{\boldsymbol{k}\propto\Bvec\times\hat{\phi}} (outward). 
The heavy-weight portions of the curves represent the locales where the
attenuation of outward emitted photons is dominated by pair creation. 
The black circles and diamonds label the colatitudes associated with 
radii equal 10\teq{\rns} and 100\teq{\rns}, respectively.}
  \label{fig:eesc_outward_parallel_ratio}
\end{figure}
\begin{figure*}
\vspace*{10pt}
\centerline{\includegraphics[width=17.5cm]{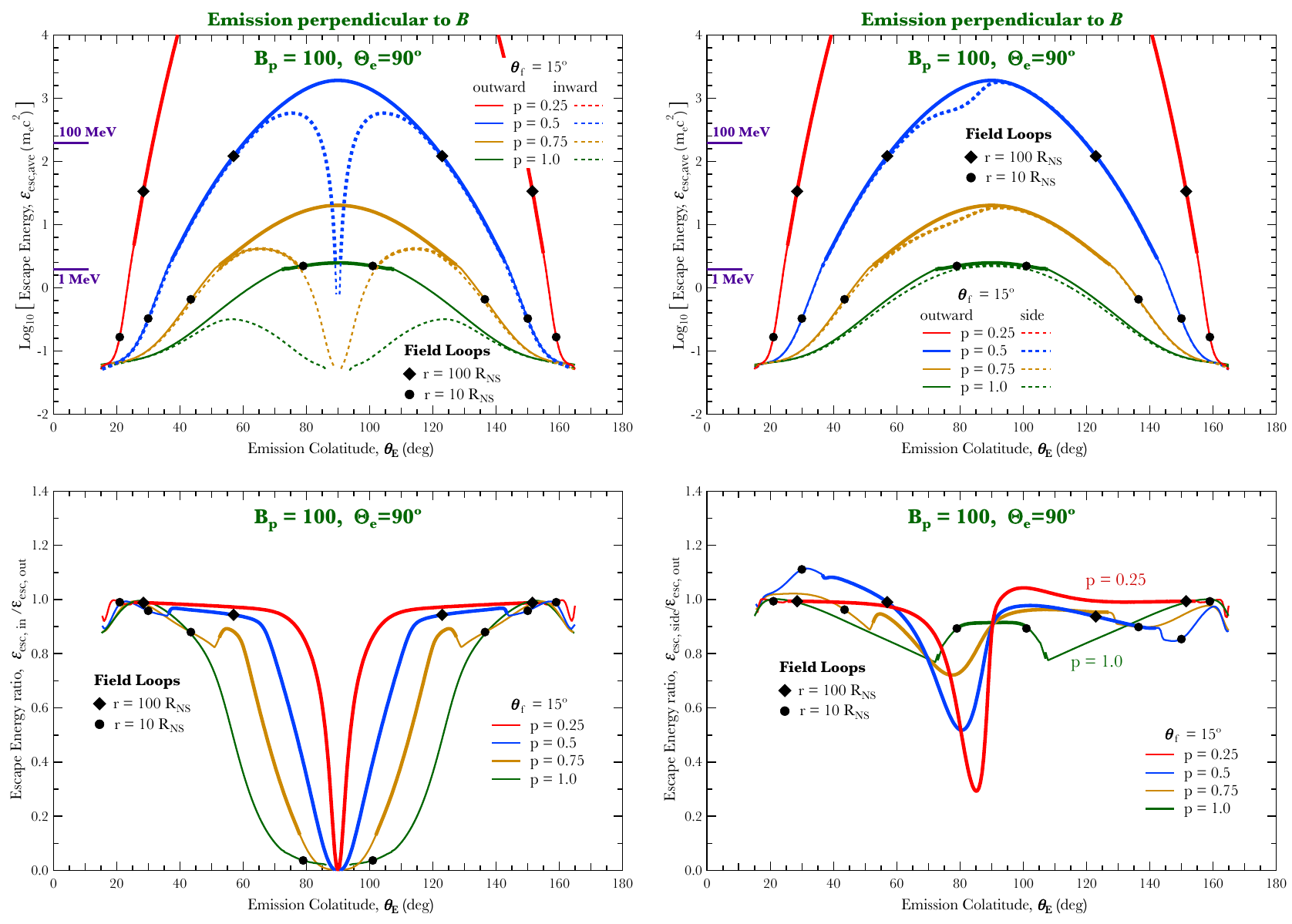}}
\vspace*{-5pt}
\caption{
Upper panels: escape energies for photon splitting and pair creation for emission 
on twisted magnetic field loops, \teq{p=0.25,\;0.5,\;0.75} and \teq{1.0}, 
ordered from top to bottom as indicated in the legends. 
Here \teq{B_p=100}, and emission is perpendicular to the local 
field (\teq{\Theta_e=90^{\circ}}).  Solid curves address 
photons generally emitted outward (\teq{\kvec\propto\Bvec\times\hat{\phi}}, giving \teq{\eescout}),
while the dashed curves therein display the escape energies for (a) inward-directed 
photons (\teq{\kvec\propto\hat{\phi}\times\Bvec}, giving \teq{\eescin}) on the left, 
and (b) sideways emission cases (\teq{\eescside} for \teq{\kvec\propto [\Bvec\times\hat{\phi}]\times\Bvec}).
The two lower panels display the ratios \teq{\eescin/\eescout} (left) and 
\teq{\eescside/\eescout} (right).   Again, heavyweight curves mark the colatitudes 
where attenuation is dominated by pair creation, and the black circles and diamonds 
mark the colatitudes where the radii equal 10\teq{\rns} and 100\teq{\rns}, respectively.}
  \label{fig:eesc_perpendicular}
\end{figure*}

To benchmark the ensuing depictions against the results displayed in
Figs.~\ref{fig:B_r_theta_phi_twist_b100} and~\ref{fig:B_r_theta_phi_twist_b10}, the
ratios of escape energies of outward emission cases to parallel emission cases are
displayed in Fig.~\ref{fig:eesc_outward_parallel_ratio} for four different
\teq{p} values and a fixed \teq{\thetaf=15^\circ}. Note the logarithmic
representation of the ratios, employed because they span around 5 decades in range.
The weighted portion of the curves marks the outward emission locales from where
photon attenuation is dominated by pair creation instead of photon splitting.
The ratio is smaller than
unity for emission colatitude \teq{\thetaE < 90^\circ} and larger than unity for
\teq{\thetaE>90^\circ}. For \teq{\thetaE<90^\circ}, at the emission local
\teq{\thetakB \equiv \Thetae =90^\circ} for outward-moving photons, and the escape
energy is markedly reduced relative to the situation where emission is locally
parallel to \teq{\Bvec}. For emission colatitudes \teq{\thetaE>90^\circ} in the
other hemisphere, parallel emitted photons travel close to the stellar surface where
the field strength is strong.   Yet the \teq{\Thetae =90^\circ} photons still
propagate outwards into weaker fields, with trajectories that are symmetric about
the magnetic equator, i.e. under the interchange \teq{\thetaE \to 180^{\circ} -
\thetaE}. The combination of these influences leads the escape energy ratio to
generally be an increasing function of \teq{\thetaE}. The deviation of the escape
energy ratio from unity at small and large colatitudes is enhanced for small
\teq{p}, primarily because the radial component of the photon momentum vector is
enhanced with larger twists for the case of photons emitted parallel to
\teq{\Bvec}. 

Fig.~\ref{fig:eesc_perpendicular} illustrates the escape energies for photons
emitted perpendicular to field loops that are characterized by footpoint colatitude
\teq{\thetaf=15^\circ} and four different \teq{p} values. In the top two panels, the
solid curves represent the outward emission case where photons possess initial
momenta \teq{\kvec\propto\Bvec\times\hat{\phi}} and propagate away from the star.
The escape energies for the inward case (\teq{\kvec\propto-\Bvec\times\hat{\phi}})
are displayed in the upper left panel as dashed curves. For both emitting
directions, the photon trajectories lie in ``meridional'' \teq{(r,\theta)} planes so
that the escape energy curves are symmetric about the magnetic equator. The outward
emission escape energy curves resemble the bell shapes in the parallel emitting
case, with ratios of the two as depicted in
Fig.~\ref{fig:eesc_outward_parallel_ratio}. For the inward case, the photon escape
energy is strongly diminished near the equator, generating a double-peaked shape.
This reduction is because the trajectories of these photons pass close to the
stellar surface; the gaps in the inward curves demarcate the locales where emitted
photons actually hit the star. These colatitudes of stellar shadowing of the
radiation get reduced as the twist increases since the equatorial portions of the
loops move to higher altitudes, for which the star becomes more remote. The cusps in
the \teq{p=1} outward emission curves near \teq{\thetaE=72^\circ} and
108\teq{^\circ} are caused by the fact that pair creation attenuation coefficient is
not a strictly increasing function of photon energy between two distinctive Landau
levels. Thus the escape energy exhibits two small steps which correspond to the
first and second Landau levels (see Appendix A).

Escape energy curves for photons emitted in one of the sideways directions
[\teq{\kvec\propto(\Bvec\times\hat{\phi})\times\Bvec}] are displayed as dashed
curves on the upper right of Fig.~\ref{fig:eesc_perpendicular}. The curves are 
now not symmetric about the equator, and the values of the escape
energy are intermediate between those of the outward and inward emission cases. The
escape energies for another side direction
[\teq{-\kvec\propto(\Bvec\times\hat{\phi})\times\Bvec}] are reflection-symmetric 
about the equator to the depicted ones, and thus they are not explicitly displayed.

To complete the suite of information,  the escape energy ratios
\teq{\varepsilon_{\rm esc,in}/\varepsilon_{\rm esc,out}} and \teq{\varepsilon_{\rm
esc,side}/\varepsilon_{\rm esc,out}} are presented in the lower panels of
Fig.~\ref{fig:eesc_perpendicular}.  This row augments the upper panel information
wherein it is difficult to discern escape energy curves that differ by 10\% or less.
The \teq{\varepsilon_{\rm esc,in}/\varepsilon_{\rm esc,out}} ratios are close to
unity for large or small emission colatitudes and drop to around zero near the
equator. As the twist increases, the emission locale raises and the ratio drops
arise at colatitudes somewhat closer to the equator. Therefore the differences
between outward and inward emission cases decrease with higher field twists. The
\teq{\varepsilon_{\rm esc,side}/\varepsilon_{\rm esc,out}} ratios also decline near
the equator. The equatorial bump in the \teq{p=1} case is caused by the onset of
pair creation in the outward emission case.  Curve structure bracketing 
the bump, with analogous variations exhibited for other \teq{p} values (albeit 
less pronounced), captures the discontinuities of the pair conversion rates 
near threshold; see Appendix A.   The shapes of the \teq{\varepsilon_{\rm
esc,side}/\varepsilon_{\rm esc,out}} ratios are generally complicated since photons emitted
sideways do not move in fixed \teq{(r,\theta)} planes. Yet the ratios are generally
close to unity (20\%) for most emission colatitudes.

In summary, the main message contained in the results depicted 
in Figs.~\ref{fig:eesc_outward_parallel_ratio} and~\ref{fig:eesc_perpendicular}
is that emitting photons at large angles to the local field alters the 
magnetospheric opacity substantially, and that the escape energy is 
most sensitive to the azimuthal direction of emission around \teq{\Bvec} 
for equatorial locales.  Both these properties emerge naturally from 
the angle and field dependence of the pair creation and photon splitting 
rates.

\subsection{Polar Emission Zones}
 \label{sec:polar}

The final focus of our results section is on the regions very close to the magnetic
poles. This is primarily motivated by the phenomena of magnetar giant flares, yet it
may also be germane \citep{Younes-2021-Nat-Ast} to the simultaneous detection of a
fast radio burst \citep[FRB;][]{Bochenek-2020-Nature} with a hard X-ray one
\citep[FRB-X;][]{Mereghetti-2020-ApJ,Ridnaia-2021-Nat-Ast} from the magnetar SGR
1935+2154 on April 28 (UTC), 2020.

Transient giant flares possess enormous luminosities \teq{10^{44} - 10^{47}} erg/sec
at hard X-ray energies, and constitute the hardest emission signal known for
magnetars. Their initial spikes, generally lasting less than around 0.2 sec, are
observed to extend up to the MeV-band energies that permit pair creation to possibly
be active. Specifically, for the Galactic magnetar giant flares,
\cite{Hurley-1999-Nature} identified emission up to around 2 MeV for the August 27,
1998 event from SGR 1900+14, while \cite{Hurley-2005-Nature} reported signals up to
around 1 MeV for the December 27, 2004 giant flare from SGR 1806-20.  Going beyond
the Milky Way, the April 15, 2020 giant flare from a magnetar in the NGC 253 galaxy
was far enough away that only the initial spike was observed, and it was not subject
to instrumental saturation influences. This event thus supplied unprecedented
time-resolved spectroscopy. {\it Fermi}-GBM observations of the transient detected
photons up to around 3 MeV \citep{Roberts-2021-Nature}, and the interpretation was
that the initial spike constituted soft gamma-ray emission from relativistic plasma
outflow from the polar (perhaps open field-line) regions of a rotating magnetar. 
This picture motivates the focal investigation of this subsection.

\begin{figure*}
\vspace*{10pt}
\centerline{
\includegraphics[width=18.0cm]{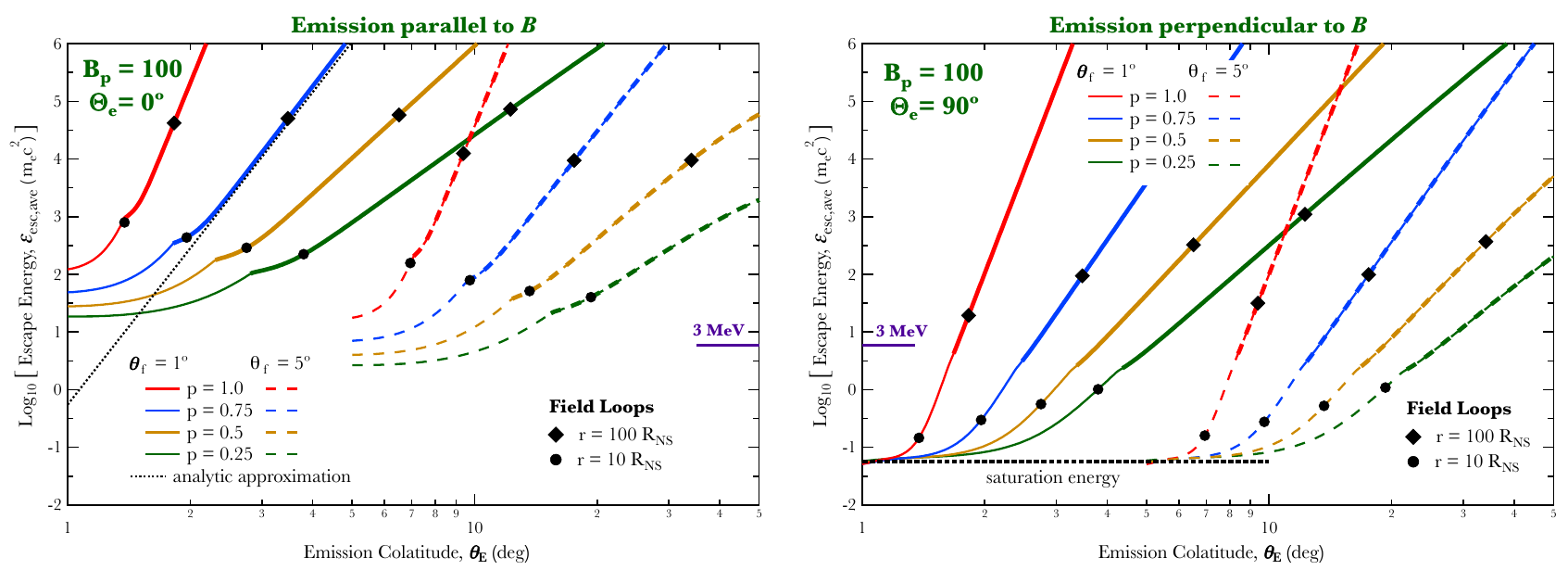}}
\vspace*{-5pt}
\caption{
Left: Polarization-averaged escape energies plotted as functions of emission colatitude \teq{\thetaE} 
(note the logarithmic scale) for photons emitted parallel to the local field directions 
(\teq{\Theta_e=0^\circ}) from field loops with footpoint colatitudes \teq{\thetaf=1^\circ} (solid) 
and \teq{\thetaf=5^\circ} (dashed) for four different \teq{p} values.  As in 
prior figures, the black circles and diamonds mark the colatitudes where the 
emission altitude equals 10 \teq{\rns} and 100 \teq{\rns}, respectively. 
The black dotted curve is the escape energy for the \teq{p=0.5} case calculated 
using the analytic formula in Eq.~(\ref{eq:eesc_analytic_approx}),
with \teq{\thetaf} 
adjusted to accommodate the increase of the field line footpoint colatitude in curved spacetime. 
Right:  Escape energies for photons emitted perpendicular to the local field directions 
(\teq{\Theta_e=90^\circ}, outward-directed, i.e., \teq{\kvec\propto\Bvec\times\hat{\phi}})
from field loops anchored at \teq{\thetaf=1^\circ} (solid) and 
\teq{\thetaf=5^\circ} (dashed). The horizontal dashed line indicates the asymptotic 
escape energy [\teq{\approx 26}keV; see Eq.~(\ref{eq:eesc_saturation}) and the related text].  }
  \label{fig:eesc_polar}
\end{figure*}

For the polar surface field \teq{B_p=100} case, Fig.~\ref{fig:eesc_polar} displays
the escape energies for photons emitted parallel to the local magnetic field
directions from field loops anchored at polar regions, specifically with
\teq{\thetaf=1^\circ} and \teq{5^\circ}. The left panel partners three of the panels
in Fig.~\ref{fig:B_r_theta_phi_twist_b100}, while the right panel is to be compared
with the solid curves in the upper panels of Fig.~\ref{fig:eesc_perpendicular}.   In
both cases, the escape energies increase with the emission colatitude \teq{\thetaE}
and realize power laws when pair production is the dominant mode of photon
attenuation, corresponding to emission altitudes larger than around 10 \teq{\rns}.
This is a domain where the GR influences are small, so that a flat-spacetime
asymptotic analysis can be used to describe the analytic character of the escape
energy. For the emission parallel to \teq{\Bvec}, the case in the left panel, this
adapts the protocol developed by \cite{SB14}. The power-law behavior can be well
estimated by an analytic approximation:
\vspace{-3pt}
\begin{equation}
   \eesc  \;\approx \; A(p) \frac{h^{2+p/2}}{B_p\thetaf} 
   \;\approx\;  A(p) \frac{\thetaE^{1+4/p}}{B_p\thetaf^{2+4/p}} \quad ,
 \label{eq:eesc_analytic_approx}
\end{equation}
where the poloidal field line relation in Eq.~(\ref{eq:h_theta}) is used in the 
second approximation, and
\begin{eqnarray}
   &A(p)& \;=\; \frac{2^5 (p+3)^{p+3}}{3 p(p+2)^{p+3}} \nonumber \\
   &&\times \left\{\log{\left[ \frac{3^2p(p+2)^{3p/2+5/2}}{2^8 (p+3)^{3p/2+4}}
   \frac{\fsc\rns}{\lambdabar_c}\right]} \right\}^{-1} \;.
\end{eqnarray}
Here \teq{h=r/\rns} is the altitude of the emission locale scaled by the stellar
radius \teq{\rns}. This relation is an extension of the Eq. (20) of \cite{SB14} to
the twisted field configuration. It is valid for arbitrary twist parameter \teq{0 <
p \leq 1} with \teq{\thetaE \ll} 1. For a specific \teq{p}, one can infer that
\teq{\eesc \propto h^{2+p/2} \propto \thetaE^{1+4/p}}, with a power-law index
that is independent of the field strength \teq{B_p} and the footpoint colatitude
\teq{\thetaf} of the field loops. The derivation of
Eq.~(\ref{eq:eesc_analytic_approx}) is detailed in Appendix B.
Eq.~(\ref{eq:eesc_analytic_approx}) is plotted for \teq{p=0.5} and
\teq{\thetaf=1^\circ} as the black dotted curve in Fig.~\ref{fig:eesc_polar} to
compare with the numerical results. Although not depicted in the figure,
Eq.~(\ref{eq:eesc_analytic_approx}) well approximates (with error \teq{\lesssim30\%}
for \teq{\thetaE\lesssim20^\circ} and \teq{h>10}) the power-law relationships for
all the other \teq{p} cases. Observe that there is a slight downward curvature at
high colatitudes beyond the \teq{\thetaE\ll 1} domain.

The right panel of Fig.~\ref{fig:eesc_polar} illustrates the escape energies for
photons emitted perpendicular (outward-directed with
\teq{\kvec\propto\Bvec\times\hat{\phi} }) to the field loops for the same polar
footpoint colatitudes. Note that the escape energies for other perpendicular
emission cases (inward-directed, sideways) are almost identical to the outward case.
 The values of the escape energies are smaller than their counterparts for the
parallel emission case, since photons immediately move across field lines, thereby
manifesting a different power-law behavior with \teq{\eesc\propto \thetaE^{2+4/p}}.
This index can be quickly obtained by inserting \teq{\thetakB=\pi/2}, \teq{B\propto
B_p/h^{p+2}}, and Eq.~(\ref{eq:h_theta}) into \teq{B\omega_{\perp}\approx} constant;
similar behavior is realized for other fixed values of \teq{\thetakB}.

The escape energy curves also realize a horizontal saturation (black dashed line) at
small emission colatitudes \teq{\thetaE} close to the stellar surface.  The saturated 
escape energy is independent of the twist parameter \teq{p}, and can be
estimated in a superstrong field where \teq{{\cal M}_1\approx1/(6B^3)} dominates 
the reaction rates \teq{{\cal M}_\sigma} (see Eqs.~\ref{eq:calM_i_form} and
\ref{eq:Lambda_s_def}) and the photon splitting coefficient is independent of the
field strength. In this case, Eq.~(\ref{eq:split_pol_ave}) yields
\begin{equation}
  \eesc\; \approx \;\sqrt{1-\PsiE}\left(\frac{1440\pi^2 \lambar_c}{\fsc^3 \rns\, \sin^6\Thetae}\right)^{1/5}\quad.
 \label{eq:eesc_saturation}
\end{equation}
Here we assume that the stellar radius \teq{\rns} is the typical lengthscale for the
attenuation. For emission locales near the surface, Eq.~(\ref{eq:eesc_saturation})
gives a saturated escape energy \teq{\eesc\approx26} keV when \teq{\Thetae = \pi/2}
and \teq{\PsiE=0.425}, which is in agreement with the values near the footpoint
colatitudes for both \teq{\theta_f=1^\circ} and \teq{5^\circ} in the right panel of
Fig.~\ref{fig:eesc_polar}. For \teq{\Thetae < \pi/3}, the typical attenuation locales
are well removed from the surface, and then Eq.~(\ref{eq:eesc_saturation})
accurately expresses the saturation energy when \teq{\PsiE\to 0} and \teq{\eesc
\approx 36\,\hbox{keV}\, (\sin\Thetae )^{-6/5}}.

A key feature of the polar region emission is that the escape energy for parallel
emitted photons increases with the decrease of the field loop footpoint colatitude
\teq{\thetaf}, for fixed field strength \teq{B_p} and twist parameter \teq{p}.
Therefore in the polar region, the escape energy can be very large (\teq{>
10-100}MeV) even for photons emitted near the surface, since the field lines are
only mildly curved. Adding twists to the field configuration will further enlarge
the escape energy because the radius of the curvature is increased.  According to
Fig.~\ref{fig:eesc_polar}, this is more pronounced for the case of emission parallel
to \teq{\Bvec}, much less so for emission orthogonal to the field.  This rise in
\teq{\eesc} as \teq{\thetaf} becomes small applies to both processes, so that the
competition between pair creation and photon attenuation in polar regions appears to
be only modestly dependent on the twist parameter \teq{p}.

The 3 MeV energy markers on both panels of Fig.~\ref{fig:eesc_polar} signify the
approximate maximum energy observed from the initial spike of the magnetar giant
flare in the NGC 253 galaxy \citep{Roberts-2021-Nature} by {\it Fermi}-GBM in 2020.
It is clear from the left panel of Fig.~\ref{fig:eesc_polar} that photon transparency
in the inner magnetosphere for such a signal is guaranteed right down to
the surface if the pertinent field line footpoint colatitude is somewhat smaller
than \teq{5^{\circ}}, and the emission is along the field.  In striking contrast, if
the emission is perpendicular to the field, and outward directed (right panel), then
photon splitting and even pair creation would be rife, precluding the visibility of
such a signal if generated at altitudes of \teq{30\rns} or less. The \teq{\eesc}
curves are very similar for inward-directed, \teq{\Thetae=\pi /2} emission (not displayed). 
Moreover, our computed saturation escape energies for photon splitting are 295 keV
for \teq{\Thetae = 10^{\circ}} and 674 keV for \teq{\Thetae = 5^{\circ}} for all
axisymmetric twist scenarios; see Eq.~(\ref{eq:eesc_saturation}). While Doppler
boosting will likely beam the hardest giant flare emission along the local field,
substantial angles \teq{\Thetae} to the field will be germane to emission at
energies below 1 MeV.  Accordingly, the time-dependent spectra presented in
\cite{Roberts-2021-Nature} likely constrains the inferred quasi-polar emission
altitude to at least \teq{10\rns} for highly-twisted (\teq{p=0.25}) configurations
and more than \teq{30\rns} for dipole morphology.  Refined emission geometry
diagnostics for this NGC 253 transient will be deferred to future work.

\section{Context and Discussion}
 \label {sec:obs_context}

To enhance the insights delivered by the escape energy results, 
the focus here is first on how opacity regions change with increases in twist, 
and then on the connections between pair creation and twists 
informed by pulsar understanding and magnetar radio emission.

\subsection{Opacity Volume and the impact of field twists} 
 \label{sec:opacity_volume}

In this subsection, an exploration of how the twisted field structure affects the
opaque volumes for photon splitting and pair creation in the magnetosphere is
presented. In contrast to Section~\ref{sec:escape_energies}, here photons are
emitted from a large variety of fixed locales instead of from individual field
loops. The twists change the morphology of the magnetic field around the star,
thereby altering the transparency of the magnetosphere.

\begin{figure*}
\vspace*{10pt}
\centerline{
\includegraphics[width=8.75cm]{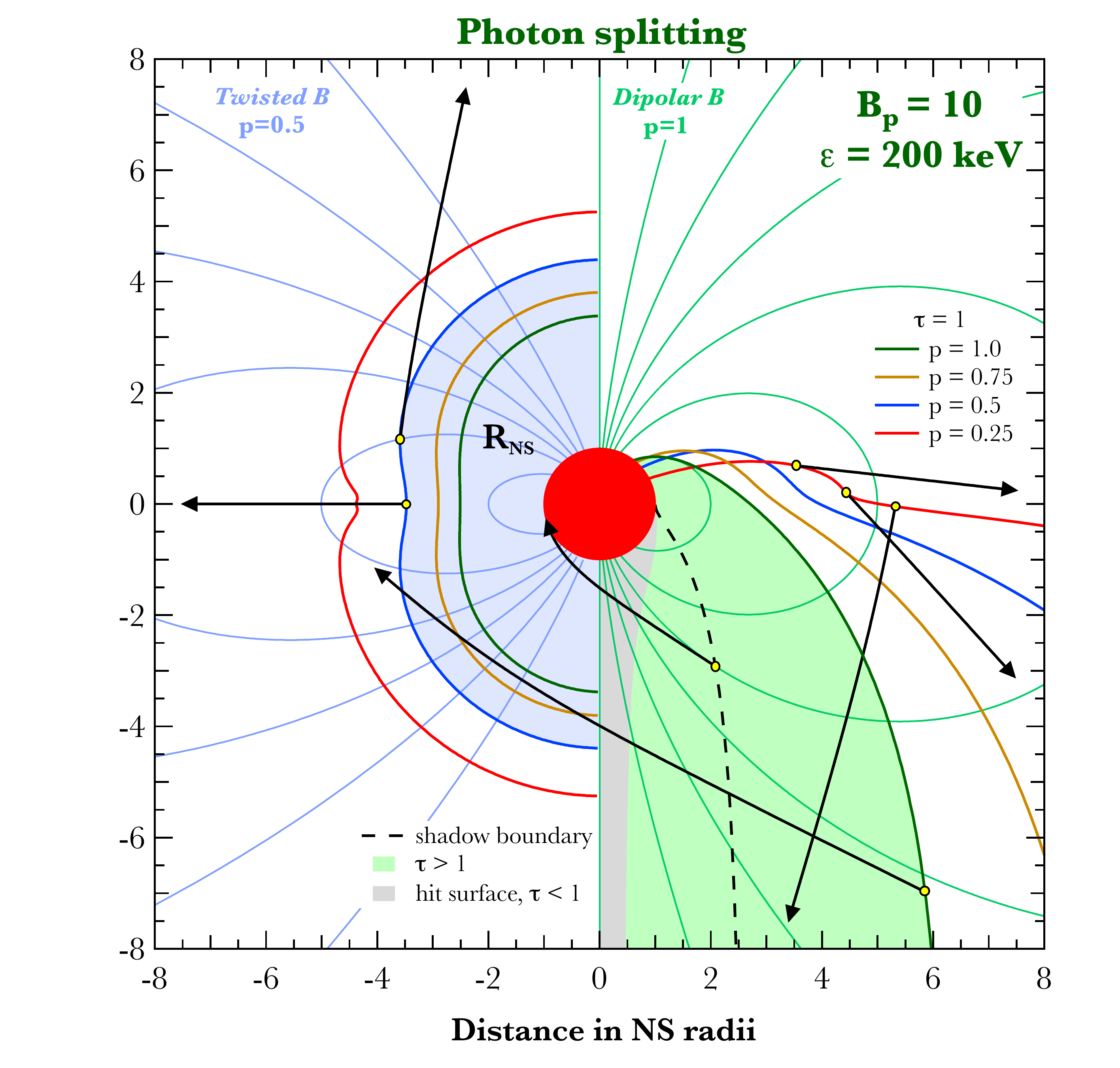}
\hspace{-10pt}
\includegraphics[width=8.75cm]{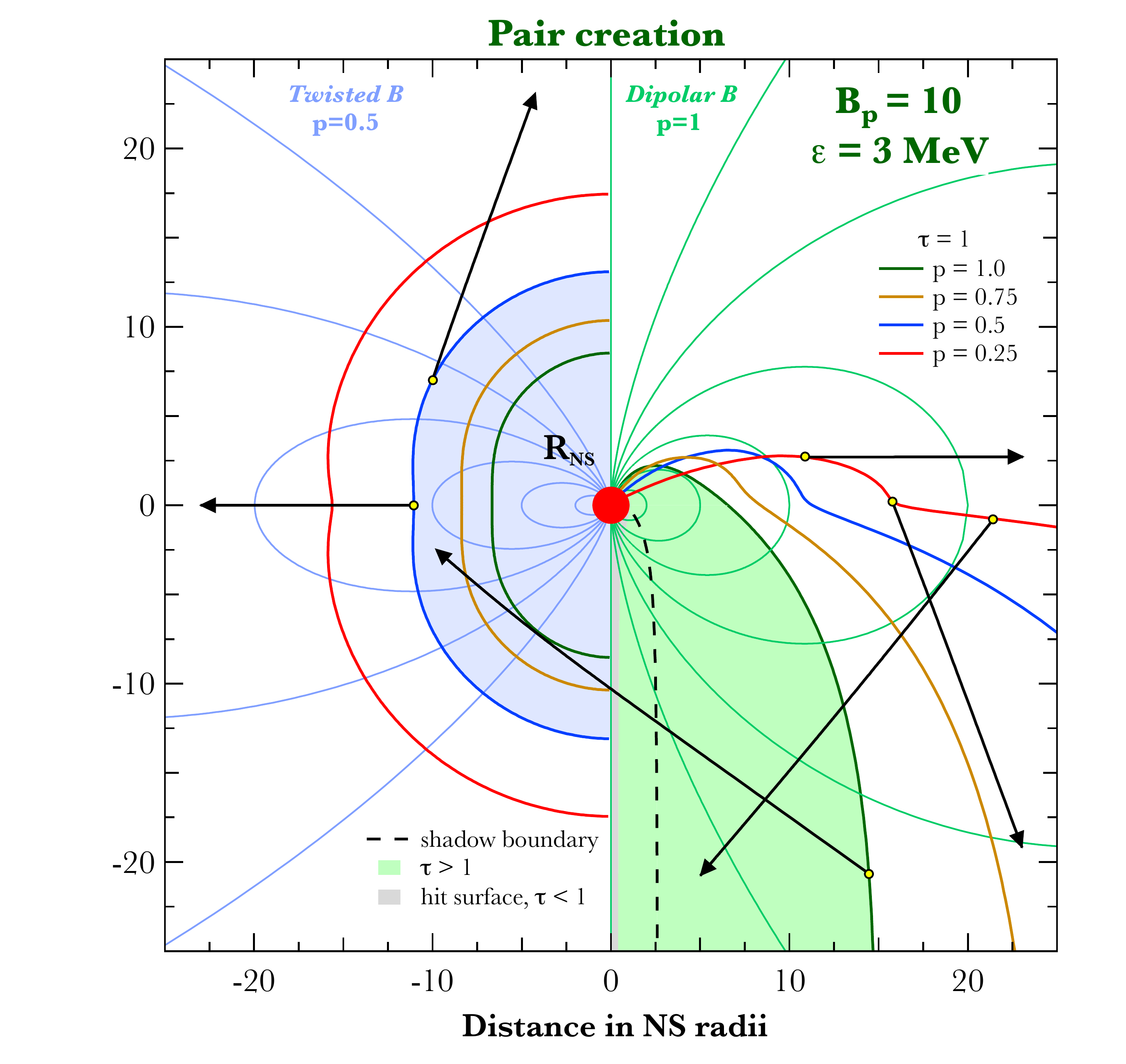}}
\vspace*{-10pt}
\caption{
Sections of opaque volumes of polarization-averaged photon splitting (left panel)
and pair creation (right panel) for photons with energy equal to 200 keV and 3 MeV,
respectively.  Photons are emitted in twisted magnetospheres in the Schwarzschild
metric with \teq{M=1.44M_{\odot}} and \teq{\rns = 10^6} cm.  In each panel, colored
contours are plotted for photons emitted outward and perpendicular to local field
directions (left half) or parallel to the field directions inside the meridional
plane (right half), with \teq{p = 0.25, 0.5, 0.75}, and \teq{1.0}, as labelled, 
and generally ordered from lower to higher altitudes.  All
photon trajectories lie inside a meridional plane containing the star's center. 
Photons emitted inside the contours are attenuated by photon splitting (left) or
pair conversion (right). The shaded regions represent the opaque volumes for the
outward emission with \teq{p=0.5} (blue) and for the parallel emission with
\teq{p=1} (green), respectively. Photons emitted in these regions will be attenuated
(\teq{\tau > 1}) before propagating to infinity or hitting the stellar surface.  The
dashed black curves define the boundaries of the shadow regions where emitted
photons are on trajectories that will hit the surface; photons emitted in the gray
interior shaded regions will hit the surface before being attenuated (\teq{\tau < 1}).  
Black curves with arrows signify photon paths for selected emission locales on
different contours.  Displayed magnetic field lines are projections onto the
meridional plane; they are plotted in light blue (left, \teq{p = 0.5}) and light
green (right, \teq{p = 1}; dipole).
 }
  \label{fig:opaque_volume}
\end{figure*}

The left panel in Fig.~\ref{fig:opaque_volume} displays the opaque regions of
polarization-averaged photon splitting for photons with energy \teq{\varepsilon=200}
keV emitted in the twisted magnetospheres with \teq{p} = 1.0, 0.75, 0.5, and 0.25.
The colored contours depict the boundaries of the opaque regions in the planar
meridional section that contains the center of the star.  Thus at the boundary, the
photon splitting escape energy is \teq{200}keV, and inside it \teq{\eesc} is lower;
these zones of opacity are signified in two cases by shading colored to pair with
that of the corresponding boundary contours.  Given the axi-symmetric field
construction, the volumes of opacity are formed by rotating these planar sections
about the magnetic axis. Field loops are plotted as projections onto the planar
meridional section for \teq{p = 0.5} on the left-hand (light blue curves) and for
\teq{p = 1.0} on the right-hand (light green curves) sides, respectively. The
maximum radii \teq{\rmax} for the field loops are fixed at 2, 5, 10, 20, 50, and 200
\teq{\rns}; accordingly, the field loops on the left-hand side (\teq{p} = 0.5)
appear flattened by the twists that generate field components out of the meridional
plane and move the footpoints closer to the equator.

Contours on the left-hand side of the panel present the opaque regions for photons
emitted perpendicular to the local field directions specifically with their
trajectories lying in the meridional plane
(\teq{\kvec\propto\Bvec\times\hat{\phi}}), therefore the contours are symmetric
about the \teq{x}-axis. For a fixed photon energy, the opaque volume increases with
the increase of the magnetospheric twist, character that is not that easily
discerned from the figures in Section~\ref{sec:escape_energies}; there photons are
emitted from field loops with fixed \teq{\thetaf} (i.e., see
Fig.~\ref{fig:eesc_perpendicular}) and the coupling of escape energies to emission
altitudes is highlighted. At the equator in Fig.~\ref{fig:opaque_volume}, emitted
photons are directed away from the star, experiencing weaker field strengths along
their trajectories compared with photons from other emission colatitudes. Therefore
the overall opacity is smaller and the contours slightly shrink near the equator.
This is more obvious for magnetospheres with large twists, e.g., \teq{p} = 0.25,
where the magnetic field direction changes more rapidly near the equator.

On the right-hand side of the panel, opaque boundary contours are displayed for
photons emitted parallel to the poloidal components of the magnetic field. This is a
twisted magnetosphere analog of Fig. 9 in \cite{Hu-2019-MNRAS}. The opaque contours
are not symmetric about the \teq{x}-axis since momentum directions are different for
photons emitted in the upper and lower hemispheres. The opaque volumes shrink near
the north magnetic pole for smaller \teq{p}. Thus the vicinity of the north pole
becomes more transparent as the twist increases. This is caused by the enhancement
of the radial field component when increasing the twists. Therefore both the optical
depth \teq{\tau} and the \teq{B \sin{\thetakB}} factor that controls it grows slowly
for photons emitted near the north pole.  This is in accordance with the
surface-emission curves in Fig.~\ref{fig:B_r_theta_phi_twist_b100}. At colatitudes 
near the equator, the opaque volume expands with a decrease in \teq{p}, 
because the \teq{B \sin{\thetakB}} factor sampled by the emitted photon receives 
significant contributions from the toroidal field component.  When the emission
colatitude \teq{\thetaE} increases across the equator, the opaque volume expands
abruptly, creating small dips on the contours above the equator. This is because
photons emitted in the lower hemisphere are directed toward the star (see the black
photon trajectories in Fig.~\ref{fig:opaque_volume}). The lower hemisphere is more
opaque since emitted photons sample stronger field strengths and shorter radii of
field curvature near periastron.

The right panel in Fig.~\ref{fig:opaque_volume} presents the opaque volumes to
photons with observed energy \teq{\varepsilon=3} MeV caused by polarization-averaged pair
creation. The shapes of the contours are very similar to their photon splitting
analogs on the left, yet the scales are larger due to the higher photon energy
required to exceed the pair threshold, something that is also apparent in
Figs.~\ref{fig:eesc_indiv_spli_pair} and \ref{fig:B_r_theta_phi_twist_b10}.  As on
the left, the transparent volume proximate to the north pole is enlarged as the
twist increases, implying reduced rates of pair creation at altitudes \teq{\lesssim 8\rns}. 
The green and blue shaded areas, bounded by the dashed black curves, 
represent the opaque regions where emitted photons will be attenuated by 
splitting or pair conversion before propagating to infinity or striking the surface.
The gray shaded area displays the interior shadow region where emitted photons 
actually reach the surface without attenuation (\teq{\tau < 1}). 
While not explicitly depicted, we remark that for photons of fixed energy
\teq{\varepsilon>2m_ec^2} emitted parallel to \teq{\Bvec} within the meridional
plane, the opaque volumes of pair creation are larger than the photon splitting ones
when the emission colatitude is not too small, \teq{\thetaE\gtrsim 60^{\circ}}.

The results displayed are for a choice of \teq{B_p=10}.  When the polar 
magnetic field is increased to \teq{B_p=100}, the opacity volumes for both splitting 
and pair creation increase somewhat, as expected: a rise in the field strength 
throughout the magnetosphere increases the rates for both processes at each locale.

\vspace{0pt}
\subsection{Discussion}
 \label{sec:discussion}

The suite of results presented so far evinces two clear trends.  First, there is a
general increase of the opacity volumes in the inner magnetosphere when the twist is
increased. This coupling is driven by the rise in the magnitude of the field, though
a notable exception is in non-equatorial zones for outward emission parallel to
\teq{\Bvec}.  The second key trend is an increase of the escape energies with larger
twists in the case of emission parallel to \teq{\Bvec} on specified field loops
somewhat near the poles. This behavior is caused by a general straightening of field
lines, which then yields predominantly higher altitudes with lower fields where
opaque conditions arise. While these trends may seem somewhat contradictory, they
are actually encapsulated in the crossing over of opacity boundaries with different
\teq{p} values in the upper hemispheres of the panels in
Fig.~\ref{fig:opaque_volume}. Addressing the impact of twists specifically on pair
creation is also germane, since pair populations are believed to underpin the
currents that establish the twisted fields on extended ranges of altitudes in the
magnetosphere.

The commentary on Figure~\ref{fig:eesc_polar} in Section~\ref{sec:polar} focused on
the gamma-ray transparency for magnetar giant flares.  Also apparent in this polar
zone figure is that \teq{\eesc} for the onset of pair creation domination of photon
splitting increases modestly as the field twist rises, when the emission is parallel to
\teq{\Bvec}.  Yet, these pair onset \teq{\eesc} values do not change appreciably
with variations in the twist parameter \teq{p} for emission orthogonal to the field.
This reasonably implies that once sufficient pair creation is established to support 
significant twists, the configuration can sustain the twist as long as the energy source 
for a giant flare continues; the twist does not unduly starve itself of pairs.  The pair 
conversion appears likely to occur at altitudes of 20-30 stellar radii, tapping the giant 
flare radiation supply there, only to cease at somewhat higher altitudes when the flare
plasma outflow \citep[likely ultra-relativistic: see][]{Roberts-2021-Nature} becomes
optically thin to splitting and \teq{\gamma\to e^{\pm}} (yet still Thomson optically
thick) and the radiation we see emerges.

The situation for persistent magnetospheric signals from magnetars may be
quite different.  Earlier figures such as Fig.~\ref{fig:B_r_theta_phi_twist_b10}
and~\ref{fig:eesc_perpendicular} concentrate on larger footpoint colatitudes that
are more or less commensurate with the active twist ones in the plasma
simulations of \cite{CB17}. For these field lines, the photon energies for the onset 
of pair creation in quasi-equatorial regions are only moderately increased
by a strengthening of the twist.  These field zones likely address the persistent hard X-ray
tail emission (Thomson optically thin) of magnetars, which is known to not extend 
up to pair threshold \citep[e.g.,][]{Kuiper06,Hartog08a,Hartog08b}.  
Specifically, photons are easily generated to energies well in excess of 1 MeV 
in resonant inverse Compton models  \citep{Wadiasingh18}, tapping the energies of 
ultrarelativistic pairs accelerated in the magnetosphere.  We anticipate that photon
splitting strongly attenuates these gamma-rays \citep{Wadiasingh2019}, more 
so at higher fields and for higher twists, and may effectively starve the
radiating plasma of pairs; this implies a possible limit to the maximal twist for normal
magnetar activity.  

Our results therefore highlight the need for a comprehensive inclusion of photon 
splitting opacity and its impact on pair creation inhibition in order to 
enhance simulations of dynamic, twisted magnetar magnetospheres, such as 
those studied in the works of \cite{Beloborodov-2009,Parfrey-2013,CB17}.
It also identifies the compelling need for greater observational sensitivity in the 
500 keV - 10 MeV band to afford more incisive probes of the shape
of magnetar soft $\gamma$-ray spectra \citep{Wadiasingh2019}.  This prospect 
could be fulfilled by a future Compton telescope such as 
NASA's COSI SMEX mission,\footnote{See {\tt https://cosi.ssl.berkeley.edu}}
under development, and 
AMEGO\footnote{See {\tt https://asd.gsfc.nasa.gov/amego/index.html}}
or its more compact AMEGO-X version \citep{Fleischhack-2022-ICRC}.

\subsubsection{Force-free MHD and Pulsar context for Twists}
 \label{sec:pulsar_context}

The employment of axi-symmetric force-free twists in this paper is a choice of simplicity and
convenience. Force-free magnetosphere solutions for a rotating magnetic dipole field
have been studied as models for rotation-powered pulsar magnetospheres over the last
two decades.  Solutions for both aligned \citep{Contopoulos-1999} and oblique
\citep{Spitkovsky-2006} rotators have shown that the required current varies over
the polar caps, ranging from super-Goldreich-Julian values, $J > J_{\rm GJ} =
\rho_{\rm GJ}\,c$ where $\rho_{\rm GJ} = \Omega B/(2\pi c)$ is the Goldreich-Julian
charge density \citep{Goldreich-1969} at the neutron star surface, to
anti-Goldreich-Julian values, $J < 0$.   The charges to support the force-free
magnetosphere are supplied by electron-positron production (above any  
electron-ion contribution), primarily in pair
cascades near the polar caps.  In order to supply the current in each part of the
polar cap that is compatible with the force-free global current, these pair cascades
must be non-stationary \citep{Timokhin-2010,Timokhin-2013}, producing bursts of
pairs followed by screening of the electric field.  So evidently, the magnetosphere
cannot be force-free everywhere since pair production in pulsars requires particle
acceleration.

The pair cascades are very localized in a small region near the polar caps since the
microphysical processes involved in pair production operate on scales smaller than a
neutron star radius.  The length over which particles are accelerated combined with
the photon mean-free path (essentially encapsulated in our calculations here) to
produce a pair determines the location of the pair formation front and the size of
the gap where force-free conditions are violated.  Beyond the gap, force-free
conditions can be established if the cascade can supply the global current.  If all
or parts of the magnetosphere become twisted (non-dipolar), there will be a
different global current configuration with a bigger twist requiring a larger current
\citep[e.g.][]{Beloborodov-2009,TLK02}.  Therefore, any realistic, persistent
magnetic twist must be in equilibrium with an adequate supply of pairs to supply the
current for the force-free assumption, and the force-free conditions cannot co-exist
with the regions of particle acceleration that are supplying the pairs and the
currents.  And since, as we have seen from dipolar pulsar magnetospheres, the
currents are not likely to be spatially uniform, the twists are not likely to be
axisymmetric as we have assumed.  Our adoption of a force-free, axisymmetric twist
is therefore idealized, and assumes that the current generated by the twist can be
supplied by the pair plasma.  As we have shown, an increasing twist decreases the
opacity for pair production which may ultimately limit the amount of static
force-free twist that is sustainable with a decreasing supply of pair plasma. 
Alternatively, a non-force-free (dissipative) twist configuration may be found that
is consistent with the available pair plasma supply.

In a dynamical situation such as a magnetar burst, giant flare or enhanced
persistent emission state (outburst), a sudden increase in twist may prevent the
generation of enough pair plasma to supply the current (see the discussion above). 
In this case, the size of a force-free twist will decrease until the current it
requires is consistent with what the pair plasma can supply.  Depending on the
charge supply, this may occur before the twist is large enough to form a resistive
current sheet and undergo large-scale reconnection \citep{Parfrey-2013}. 
Accordingly, it is apparent that the intricate interplay between twist morphology,
pair creation and other sources of radiation opacity (principally photon splitting)
is an essential ingredient of next-generation modeling of force-free or dissipative
magnetar magnetospheres.

\subsubsection{Magnetar Radio Emission Connections}

This intimate interplay between field morphology and pair creation and 
photon splitting opacity is likely central to controlling the intensity and 
characteristics of persistent and transient radio emission from magnetars.  
The pair creation and cascading that precipitates such signals is thought to be
initiated by curvature radiation gamma rays in pulsars, emission that emanates from
polar field zones and is aligned closely to the local field direction
\teq{\hat{\Bvec}}. The two panels in Fig.~\ref{fig:eesc_polar} highlight the extreme
sensitivity of the escape energy to the axi-symmetric twist parameter \teq{p}, with
a dependence on the field footpoint colatitude \teq{\thetaf} as the 6th power for
the untwisted \teq{p=1} dipole, and even larger for twisted solutions; see
Eq.~(\ref{eq:eesc_analytic_approx}). In more realistic magnetosphere constructions
(i.e., more localized twists and non-ideal MHD conditions), plasma asymmetries that
couple to the stellar rotation can have an impact in spite of magnetars being slow
rotators with small polar caps and low Goldreich-Julian charge densities. The vector
vorticity of the twist relative to the rotation vector \teq{\Omegavec}
influences the polar field curvature considerably.  Associated differences in the
radius of field curvature will be reflected in the curvature emission photon energy,
which along with the field morphology leads to a strong sensitivity of pair creation
opacity to polar field geometry.

Thus small changes in the local field and accompanying modifications to the locale
and shape of the pair formation front across the polar cap should have a dramatic
impact on pulsed radio emission, its efficiency and spectrum, and also the beam
morphology. Sensitive observations that exhibit dramatic changes in radio signals 
\citep[e.g.][ for Swift J1818]{Lower-2021} suggest that even relatively quiescent X-ray
magnetars possess dynamic magnetospheres near their polar caps.

For more energetic transient radio emission, i.e., FRBs, charge starvation at low
altitudes near polar caps is also salient. Recently, \cite{Li-2022} reported a $\sim
35-40$ Hz quasi-periodic oscillation in the X-ray burst (FRB-X) associated with the
FRB-like radio bursts in SGR 1935+2154; this period matches well the temporal
separation of the two radio burst peaks. Such a frequency is compatible with crustal
torsional eigenmodes of neutron stars, conforming to predictions of ``low-twist''
magnetar FRB models of \cite{Wadiasingh-2019,Wadiasingh-2020} that invoke crustal
disturbances. \cite{Younes-2021-Nat-Ast} argued that the FRB-X burst of SGR 1935+2154
is of quasi-polar origin due to its unusual spectral extension.

Burst-associated plasma waves set off by the crustal disturbances 
can trigger low-altitude pair cascades that can generate coherent
plasma oscillations and radio emission. In such situations, \cite{Wadiasingh-2020}
showed that curvature radiation photons regulate pair formation and gaps (of
voltages of the order of a TeV) in magnetars, and that splitting likely would not quench the pair
cascades.  Large persistent twists drive up pair opacity escape energies
(see Fig.~\ref{fig:eesc_polar}), and likely precipitate lower frequency curvature 
emission in the straighter field lines, thereby markedly reducing pair yields if 
the curvature photon energy is close to or below the pair escape energy.  
Accordingly, lower twists that permit modest 
pair creation at low altitudes are generally preferred 
conditions for FRB production in magnetars.

\section{Conclusion}
 \label{sec:conclusion}

In this paper, photon opacities for the processes of photon splitting and pair
creation are calculated in the twisted magnetospheres of magnetars. Fixing the polar
field strength, axisymmetric MHD twisted magnetic fields embedded in the
Schwarzschild metric were treated, for a variety of radial field scaling
parameters \teq{p} (or equivalently, the maximal twist angle 
\teq{\Delta\phi_{\rm tw}}).  Given these assumptions, adding twists to the
magnetic fields introduces toroidal field components, enhances the overall
\teq{\vert\Bvec\vert}, and straightens the poloidal field lines. The twists also
stretch magnetic field loops to higher altitudes due to the straightening of the
field lines.

The impact of twists on photon opacity and escape energies \teq{\eesc} depends on
the competition between field line straightening (decreases opacity) and field
magnitude enhancement (increases opacity). Section~\ref{sec:escape_energies}
presented escape energies \teq{\eesc} for photons emitted from specified field
loops. For photons that are emitted parallel to \teq{\Bvec},
the escape energy \teq{\eesc} rises with an increase of
the twist (lower \teq{p}), due to the straightening of the field lines 
if photons originate near the field loop footpoint, or lower fields encountered 
for emission nearer the equatorial apex of the loop. The opaque volume for
magnetospheric emission generally increases for larger twists (see
Fig.~\ref{fig:opaque_volume}), mostly because of the increase of the field
magnitude. An exception to this arises when photons are emitted parallel to the
field lines in the polar regions, where field line straightening overrides the
impact of the enhancement of the field magnitude.

In Section~\ref{sec:polar}, which focused on photon opacity in the polar regions,
the escape energy \teq{\eesc} of photons emitted parallel to \teq{\Bvec} increases
for larger twists and smaller footpoint colatitudes for the emission zones. Moreover,
adding twists generally increases the pair creation opacity volume for photons with
energies above the absolute pair threshold. Inside this opacity volume, pair
creation prevails over photon splitting and dominates photon attenuation. Again, an
exception occurs when photons are emitted parallel to \teq{\Bvec} in the polar
region, in which case the straightening of the field lines suppresses the
development of \teq{\erg\sin\thetakB} during photon propagation.
For photons emitted below pair threshold, the opaque volumes 
(now due to photon splitting) also increase.

The change of photon opacities due to the inclusion of twists has the potential to
significantly modify the spectral characters of both persistent hard X-ray emission
and magnetar giant flares. Our calculation provides a diagnostic tool to constrain
the emission geometry of both signals, pertinent to data from missions 
such as {\sl Fermi}-GBM, and could also be leveraged by future telescopes
in the MeV band such as COSI and AMEGO.

\vspace{15pt}
\centerline{\uppercase{acknowledgments}}
\vspace{-5pt}
\begin{acknowledgements}
M.~G.~B. thanks NASA for supporting this project through the
{\it Fermi} Guest Investigator Program grant 80NSSC21K1918.
Contributions from Z. W. are based upon work supported by 
NASA under award number 80GSFC21M0002.  This research has 
made use of NASA's Astrophysics Data System.
\end{acknowledgements}

\appendix

\vspace{-15pt}
\begin{flushleft}
\bf{Appendix A: Magnetic Pair Creation Rate Functions}
\end{flushleft}
 \vspace{-15pt}
\setcounter{equation}{20}
The complexity of the algebraic structure of the magnetic pair creation rate \citep{DH83}
demands a simplified approach to employing it in numerical applications to neutron star magnetospheres.
The path for this was identified in the studies of \cite{HBG97} and \cite{BH01}, and 
the hybrid exact rate+ asymptotic approximation to cover the entire phase above 
threshold was adopted in the opacity studies of \cite{SB14} and \cite{Hu-2019-MNRAS}.
We follow this protocol in this paper. 

In the energy range \teq{\erg_{\perp}\equiv \erg\sin\thetakB > 2}
just above pair threshold, the exact polarization-dependent
expressions for the production of pairs in the ground and first excited Landau state 
are employed.  Thus the coefficients for the two polarization modes 
to be employed in Eq.~(\ref{eq:pp_general}) are given by
\begin{eqnarray}
   {\cal F}^{\rm pp}_{\parallel} \; =\; \dover{2}{\erg_{\perp}^2  \vert p_{\hbox{\sevenrm 00}}\vert}
   \,\exp\left(-\dover{\erg_{\perp}^2}{2B}\right)
   & \quad , \quad & 2 < \erg_{\perp} < 1+\sqrt{1+2B} \quad , \nonumber\\[-5.5pt]
 \label{eq:calFpp_par_perp} \\[-5.5pt]
   {\cal F}^{\rm pp}_{\perp} \; =\; \dover{2E_0(E_0+E_1)}{\erg_{\perp}^2  \vert p_{\hbox{\sevenrm 01}}\vert}
   \,\exp\left(-\dover{\erg_{\perp}^2}{2B}\right)
   & \quad , \quad & 1 + \sqrt{1+ 2B} < \erg_{\perp} < 1 + \sqrt{1+ 4B} \quad .\nonumber
\end{eqnarray}
For these results, in the \teq{\parallel} case, the pairs are generated in only the 
ground state, whereas for the \teq{\perp} mode, the first excited state is 
sampled for one member of the electron-positron pair.  In these expressions,
the energies \teq{E_n} of the produced leptons and the corresponding 
momentum components \teq{p_{jk}} parallel to the magnetic field are
\begin{equation}
   E_0  = (1 + p_{01}^2)^{1/2}\quad , \quad  
   E_1  = (1 + p_{01}^2 + 2B)^{1/2}\quad , \quad
   \vert p_{jk}\vert  = \left[ \dover{\erg_{\perp}^2}{4} - 1 - (j+k)B + 
   \left( \dover{(j-k)B}{\erg_{\perp}} \right)^2\right]^{1/2}  \quad .
\end{equation}
For energies above the lowest Landau levels, it is inefficient to use exact 
expressions that sum over many terms, so we then use the asymptotic expressions 
presented by \cite{BK07} that average over the many resonances.  These forms are
\begin{equation}
   {\cal F}_{\parallel} \left(\erg_\perp,\, B\right) \;=\;
                \dover{1}{\sqrt{ {\cal L}(\erg_{\perp})\, \phi (\erg_{\perp}) }}
               \, \exp \left\{ -\dover{\phi(\erg_\perp)}{2B}  \right\}\quad ,\quad
   {\cal F}_{\perp} \left(\erg_\perp,\, B\right) \; =\;
    \dover{\erg_\perp^2-4}{2\erg_\perp^2} \, {\cal F}_{\parallel} \left(\erg_\perp,\, B\right) \quad ,
 \label{eq:BK07_perp_asymp}
\end{equation}
where \teq{\phi (\erg_{\perp}) \; =\; 2\erg_{\perp} -{\cal L}(\erg_{\perp})} and
\begin{equation}
    {\cal L}(\erg_{\perp}) \; =\;  \dover{ \erg_{\perp}^2 - 4 }{2} \,  
    \log_e\left( \dover{\erg_{\perp}+2}{\erg_{\perp}-2}\right) \quad .
 \label{eq:calL_def}
\end{equation}
Observe that all the \teq{{\cal F}_{\perp,\parallel}} functions are 
invariant under Lorentz transformations along {\bf B}, for which 
\teq{\erg_{\perp}} is constant.

%%%%

\begin{flushleft}
\bf{Appendix B: Analytic Approximation of Escape Energy at Polar Regions}
\end{flushleft}

This Appendix details the derivation of the power law analytic approximation to 
the pair creation escape energy that is illustrated in the left panel of 
Fig.~\ref{fig:eesc_polar}, and serves as a check on the numerical solutions in the quasi-polar domain.
Given that it is applicable to high magnetospheric altitudes, GR modifications can be ignored.
In the polar region, the flat-spacetime field structure in Eq.~(\ref{eq:b_twist_fl}) 
can be well approximated by a first-order series expansion in small \teq{\theta}, namely
\begin{equation}
   \Bvec\;\approx \; \frac{B_p}{r^{p+2}}\left (1,\;\frac{p}{2}\theta,
   \;\frac{1}{2}\sqrt{\frac{Cp}{p+1}}\theta^{1+2/p}\right)
   \quad ,\quad \theta\;\ll\; 1\quad.
 \label{eq:Btwist_polar}
\end{equation}
Here \teq{B_p} is the surface field strength at the magnetic poles, \teq{\theta} is the 
magnetic colatitude, and \teq{C=C(p)<1} is the twist constant in Eq.~(\ref{eqn:G_S_eqn}). 
The toroidal field component \teq{B_{\phi}} can be neglected for small \teq{\theta} since 
it is much smaller than the poloidal components.
Then the equation of a field line anchored near the magnetic pole at \teq{\thetaf} can be 
obtained by integrating \teq{dr/d\theta} in Eq.~(\ref{eq:field_loci}) in flat spacetime; this yields
\begin{equation}
   h \;=\; r/\rns\;\approx\;(\theta/\thetaf)^{2/p} \quad .
 \label{eq:h_theta}
\end{equation}
The trajectory of a photon emitted parallel to the local field direction lies essentially 
in the \teq{(r,\theta)} plane, a significant simplification for the purpose of opacity computations.
In Section 3.1 of \cite{SB14}, the optical depth for the photon emitted near the 
magnetic axis in a dipole field was analytically evaluated by integrating the 
attenuation coefficient in the meridional \teq{(r, \theta)} plane:
\begin{equation}
   \tau(s) \; =\; \frac{\fsc}{\lambar_c} \int^{\eta(l)}_0 B \sin{\thetakB} \;
   {\cal F}(\omega_{\perp},B) \frac{ds}{d\eta}d\eta.
 \label{eq:tau_integral_merid}
\end{equation}
Here \teq{s} is the pathlength along the flat spacetime straight-line trajectory, 
and \teq{\eta} is the angle between \teq{\rEvec}, the position vector of the 
emission locale, and \teq{\rvec}, the position vector of the photon along its path
\citep[see Fig. 1 of][]{SB14}.  The optical depth in twisted field configurations can be obtained 
using the same method, with Eqs. (10) and (15) of \cite{SB14} being replaced by 
\begin{equation}
   {\deltaE}\;\approx\;\frac{p}{2}{\thetaE}
   \quad , \quad 
   B \;\approx \;\frac{B_p(\deltaE-\eta)^{p+2}}{\deltaE^{p+2} h^{p+2}}
   \quad \text{and}\quad
   \sin{\thetakB}\approx\frac{p+2}{2} \eta \quad .
 \label{eq:deltaE_B_thetakB_merid}
\end{equation}
Here, \teq{\deltaE} is the angle between the photon momentum \teq{\mathbf{k}},
a constant during propagation,
and the radial direction \teq{\rEvec} at the emission locale.  Using the method 
of steepest descents to evaluate the optical depth integral in Eq.~(\ref{eq:opt_depth}), 
analogous to the protocol adopted in \cite{SB14}, the result is
\begin{equation}
   \tau_{\rm Erber}\;\approx\;\frac{3^2p(p+2)^{3p/2+5/2}}{2^8 (p+3)^{3p/2+4}}\left[\frac{p(2+p)\pi}{2 }
   \frac{\varepsilon\thetaf^3 B_p^3}{h^{4+3p/2}} \right]^{1/2} 
   \frac{\fsc\rns}{\lambdabar_c}\exp{\left[-\frac{2^5  (p+3)^{p+3}}{3p (p+2)^{p+3}} 
   \frac{h^{2+p/2}}{\varepsilon B_p\thetaf} \right]} \quad .
 \label{eq:tau_erb_approx}
\end{equation}
The polarization-averaged pair creation rate presented in Eq.~(3.3) of \cite{Erber66} 
was employed in developing this result, a rate that can be deduced by taking the 
\teq{\erg_{\perp}\gg 1} limit of Eq. (\ref{eq:BK07_perp_asymp}).
The escape energy \teq{\erg\to \eesc} can be then obtained by logarithmically 
inverting \teq{\tau_{\rm Erber}=1}, yielding
\begin{equation}
   \eesc \; \approx \;\frac{2^5 (p+3)^{p+3}}{3p ((p+2)^{p+3}} \frac{h^{2+p/2}}{B_p\thetaf} 
   \left\{\log{\left[ \frac{3^2 p(p+2)^{3p/2+5/2}}{2^8 (p+3)^{3p/2+4}} \frac{\fsc\rns}{\lambdabar_c}\right]}   
   + \frac{1}{2} \log{\left[\frac{p(2+p)\pi}{2}  \frac{\eesc\thetaf^3 B_p^3}{h^{4+3p/2}} \right]} \right\}^{-1}\quad .
 \label{eq:eesc_polar_pair}
\end{equation}
The second term in the curly brackets is much smaller than the first term 
for the range of altitudes, colatitudes and photon energies of relevance, 
so that neglecting it leads to Eq.~(\ref{eq:eesc_analytic_approx}), the
result employed in the left panel of Fig.~\ref{fig:eesc_polar}.

\newcommand{\vol}[2]{$\,$\rm #1\rm , #2.}                 

\bibliographystyle{aasjournal} 
%\bibliography{bibfile}

\end{document}